# Differential metabolic and cellular brain regional susceptibility in adult rats with chronic liver disease

**Dunja Simicic[1,2]\*, Katarzyna Pierzchala[1,2]\*, Olivier Braissant[3], Dario Sessa[4], Valérie A. McLin[4], Cristina Cudalbu[1,2]**

[1]CIBM Center for Biomedical Imaging, Lausanne, Switzerland.

[2]Animal Imaging and Technology, École polytechnique fédérale de Lausanne (EPFL), Lausanne, Switzerland.

[3]Service of Clinical Chemistry, University of Lausanne and Lausanne University Hospital, Lausanne, Switzerland.

[4]Swiss Pediatric Liver Center, Pediatric Gastroenterology, Hepatology and Nutrition Unit, Department of Pediatrics, Gynecology and Obstetrics, University of Geneva, Geneva, Switzerland.

**\*equal contribution**

**Electronic word count:** 6851 (including the abstract, references, tables, and figure legends)

**Corresponding authors: Cristina Cudalbu,** cristina.cudalbu@epfl.ch, EPFL AVP-CP CIBM-AIT Station 6 CH F0-628, Phone: +41- 21-693-7685, **Dunja Simicic**, dsimici1@jh.edu

**Number of figures and tables:** 8 figures

**Conflict of interest statements:** The authors declare no conflict of interest.

**Financial support statement**: The author(s) declare that financial support was received for the research, authorship, and/or publication of this article. DS, CC, KP were supported by the CIBM

**Authors contributions:**

DS: acquired MRS data, analyzed and interpreted MRS data, statistical analysis, creation of the first draft of the manuscript and final version and submitting the manuscript for review.

KP: performed sample collection, designed and performed histological measures, analyzed and interpreted histological data, statistical analysis of histological data, interpreted MRS data, creation of the first draft of the manuscript and final version.

DS: performed BDL surgery, follow-up of the animals and sample collection, performed histological measures, final approval of the version to be published.

OB: conceptualization, funding acquisition, interpretation of the data, drafting and reviewing of the manuscript.

VM: conceptualization, funding acquisition, interpretation of the data, drafting and reviewing of the manuscript.

CC: conceived and designed the study, analyzed and interpreted MRS data, creation of the first draft of the manuscript and final version, funding acquisition, project administration and supervision.

**Impact and implications**

Little is known about the molecular and cellular differences across brain regions during chronic liver disease, which would advance our understanding of the wide clinical spectrum of hepatic encephalopathy. We showed that while brain regions share common metabolic and cellular responses such as acute glutamine increase, glial activation and neuronal alterations, they also display unique characteristics in an established model of type C HE. The most striking regional difference was in the cerebellum, where the highest glutamine load was associated with elevated lactate, and reduced Purkinje soma surface, while the striatum showed the strongest osmolyte decrease despite having the lowest glutamine increase among the studied regions. Based on its early rise across regions, this study highlights glutamine as one of the molecular drivers of a cascade of metabolic and cellular alterations.

**Highlights:**

- Brain regions show both common and distinct metabolic and cellular responses in CLD
- Acute Gln rise across regions contributing to cellular changes
- Distinct and common glial and neuronal alterations in all brain regions
- Cerebellum was characterized by highest Gln, increased Lac and reduced Purkinje soma
- Striatum had a stronger osmolyte drop, lower Gln rise, and milder astrocytic changes

**Differential metabolic and cellular brain regional vulnerability in adult BDL rats**

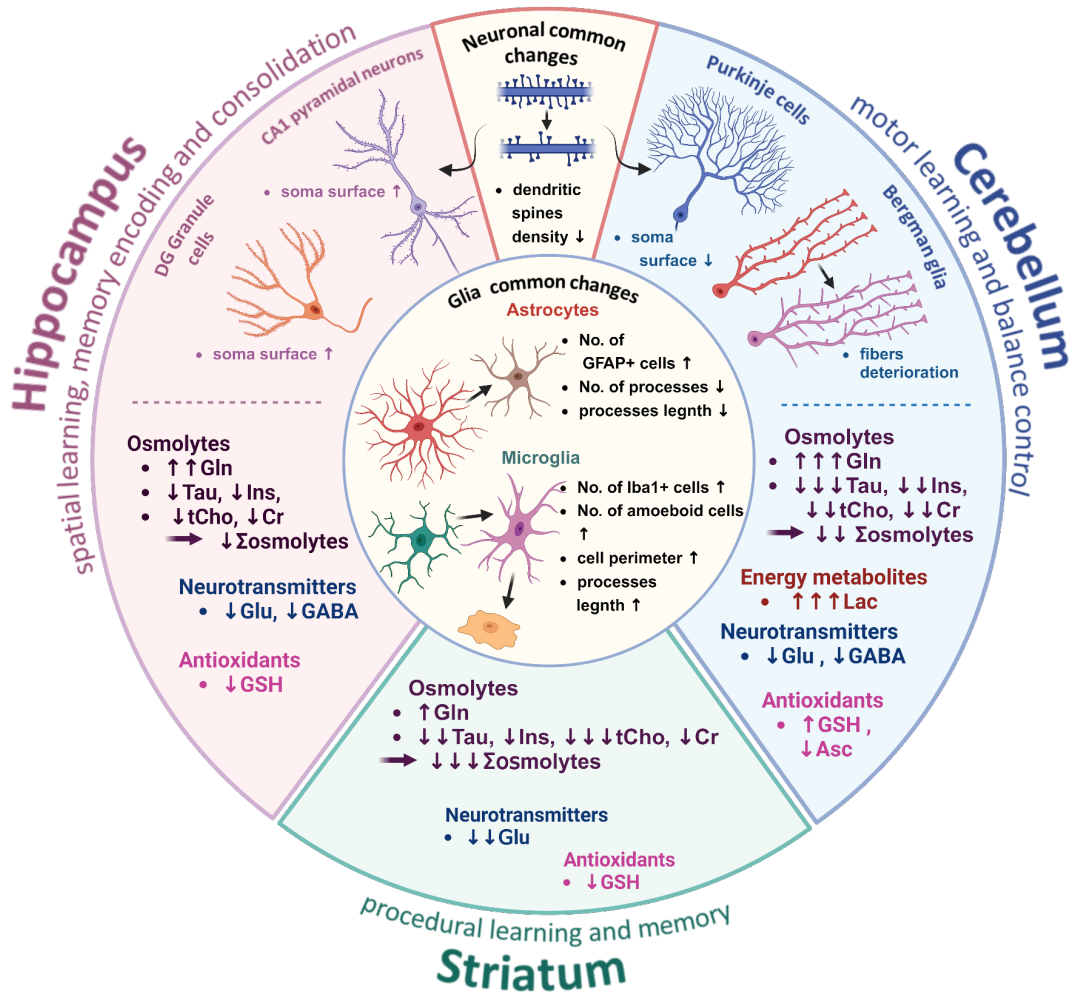

Summary of the main neurometabolic and cellular findings observed in the hippocampus, cerebellum and striatum of the BDL rat model of type C hepatic encephalopathy.

## Abstract

Background & Aims: Patients with type C hepatic encephalopathy (HE) present diverse symptoms indicating that various brain regions are affected. Understanding the distinct metabolic and cellular changes across these regions could help explain the clinical variability of HE. We analyzed, for the first time, the longitudinal *in-vivo* neurometabolic and morphological changes in astrocytes, neurons and microglia in the hippocampus, striatum and cerebellum during the progression of CLD-induced type C HE in the bile duct ligated (BDL) rat model.

Methods: Wistar rats underwent BDL surgery and their brains were studied before BDL and at post-operative weeks-2, 4, 6 and 8 (n = 45) using *in-vivo* $^1$H-MRS (9.4T) of the hippocampus, striatum and cerebellum along with histological assessment (astrocytes, microglia, and neurons) and blood biochemistry.

Results: Across all brain regions, glutamine (Gln) was the first metabolite to increase, followed by a decrease in osmolytes and neurotransmitters. The cerebellum was characterized by the highest Gln burden (+134%), elevated Lactate (+84%) and decreased GABA (-23%), suggesting its increased vulnerability. This coincided with pronounced alterations in astrocyte morphology, possibly related to the high Gln load. In contrast, the striatum displayed the lowest Gln increase (+48%) but the strongest osmolyte decrease, with milder astrocytic changes. We also showed mutual and distinct regional morphological changes in microglia (activation) and neurons (decreased dendritic spine density, soma size alterations).

Conclusions: Our findings highlight both common and differential metabolic and cellular responses to CLD across the brain regions, with cerebellum and striatum showing differential responses. We also hypothesized that glutamine is one of the initial brain metabolic markers of type C HE, impacting astrocytes and inevitably neurons contributing to cognitive decline.

**Abbreviations:**

| | | | |
|---|---|---|---|
| ALT/GPT | alanine aminotransferase | Glx | glutamine+glutamate |
| Asc | Ascorbate | GSH | Glutathione |
| Asp | Aspartate | IHC | Immunohistochemistry |
| AST/GOT | aspartate aminotransferase | Ins | Inositol |
| BDL | bile duct ligation | Lac | Lactate |
| CLD | chronic liver disease | MRS | magnetic resonance spectroscopy |
| CNS | central nervous system | NAA | N-acetyl aspartate |
| Cr | Creatine | type C HE | type C hepatic encephalopathy |
| GABA | g-aminobutyric acid | OS | oxidative stress |
| Glc | Glucose | ROS | reactive oxygen species |
| Gln | Glutamine | Tau | taurine |
| Glu | Glutamate | tCho | total choline |

## Introduction

Patients with type C hepatic encephalopathy (HE) present a wide spectrum of symptoms including motor and cognition impairment, suggesting that different brain regions may be involved and underlie the varying phenotypes[1]. Although the hippocampus, cerebellum and striatum have been identified as key regions implicated in HE in rat models and human subjects[1], little is known about the metabolic and cellular differences across these regions, something which would help understand the breadth of the clinical spectrum.

The hippocampus contributes to spatial cognition, memory and learning[2] which can be impaired by alterations in cellular morphology, neuroinflammation[3] or structural and functional connectivity[3–6]. The cerebellum plays a role in motor ability, spatial navigation and working memory[2]. Studies in animal models have suggested that neuroinflammation might affect the cerebellum earlier than other brain regions, contributing to cognitive and motor alterations[3]. Furthermore, in patients with different grades of liver disease cerebellar inflammation was demonstrated to occur very early, even before cirrhosis[5]. Among other functions, the striatum is involved in motor control, and in the reward and executive systems[2]. Cirrhotic patients show disturbed thalamostriatal connectivity, suggesting that the thalamostriatal system contributes to the cognitive deficits[7] observed in HE.

To date, few $^1$H-MRS studies in human patients with type C HE evaluated brain regional differences in metabolic ratios of tCho (total choline), Glx (glutamine+glutamate) and NAA (N-acetylaspartate) to creatine (Cr)[8], showing lower Cho/Cr in occipital cortex and higher Glx/Cr in basal ganglia. At the cellular level, astrocytes have been repetitively shown to be implicated in HE, and although emerging data suggest that other cell types, such as neurons and microglia, are also involved, detailed studies are lacking[9–11]. Thus, analyzing the different brain regions at both the cellular and molecular level in a model of type C HE might offer insights into the mechanisms underlying the complex picture of this condition[8]. In the bile duct-ligated (BDL) rat model of type C HE, we have previously shown distinct alterations in Gln between the hippocampus and cerebellum together with slight regional differences in oxidative stress and neuroinflammation[3]. These preliminary findings point to the necessity of regional brain exploration in type C HE.

Therefore, our aim was to analyze, for the first time, the longitudinal *in-vivo* neurometabolic and morphological changes of astrocytes, neurons and microglia in hippocampus, striatum and cerebellum during the progression of CLD-induced type C HE in the BDL rat model, using a multimodal approach combining $^1$H-MRS with histology and blood biochemistry. Our objective was to: 1) identify common and differential neurometabolic and cellular patterns across brain structures in a model of type C HE, and 2) pinpoint potential local or systemic drivers of disease.

1. Materials and Methods

### BDL rat model of CLD induced-type C HE

All experiments were approved by the Committee on Animal Experimentation for the Canton de Vaud, Switzerland (VD30 2/VD2439) and were conform to ARRIVE guidelines. Wistar male rats (n=45, 175-200g, Charles River Laboratories, L'Arbresle, France) were used: 35 rats underwent BDL surgery[12] (https://zenodo.org/records/10652104), while the rest were sham operated. The Supplementary Material & methods contains a detailed description of the Materials and Methods section together with information about the number of animals used for each experiment (**Table S1-S3**).

### Biochemical measurements to validate the CLD

Liver parameters (plasma bilirubin, aspartate aminotransferase (AST/GOT) and alanine aminotransferase (ALT/GPT)), glucose and blood $NH_4^+$ were measured longitudinally until week-8 post-BDL (Supplementary Material & methods, **Table S1 and Fig. S3**).

### *In-vivo* metabolic changes using [1]H-MRS

[1]H-MRS spectra were acquired on the 9.4 T system, as previously described[9], before (week-0) and at weeks-2, 4, 6, 8 post-BDL, each animal being its own control. Three different volumes of interest were selected in hippocampus ($2x2.8x2mm^3$), cerebellum ($2.5x2.5x2.5mm^3$) and striatum ($2.5x2x2.5mm^3$), with 8 to 23 rats/week scanned. LCModel was used for quantifying 18 metabolites.

### Cellular morphological changes using histology

**Immunohistochemistry (IHC):** Animals were sacrificed for histological evaluation at week 4 and 8 (n=3 per time point, n=3 sham). Astrocyte morphology was analyzed using GFAP as previously described[9], microglia morphology with Iba1 and nuclei with DAPI. Morphometric Sholl analysis was performed[13]: ~200 protoplasmic astrocytes and ~50 microglia of each group of BDL and sham rats were randomly traced for all processes revealed by GFAP and Iba1 staining respectively, in hippocampus, cerebellum and striatum (7 slides/hemisphere).

**Golgi Cox staining** was applied at 8-weeks post-BDL/sham (n=7 BDL, n=6 sham) to analyze the neuronal cytoarchitecture[14]. Only tissues uniformly stained were used for quantitative analysis (25 slides/hemisphere).

## Statistical analysis

The results of [1]H-MRS (metabolite concentrations) are presented in three different ways: bar plots, scatter plots and 1$^{st}$ axis of STATIS analysis. Given the complexity and number of the variables, we opted for a STATIS multi-table principal component analysis (RStudio, Supplementary Material & methods)[15] to evaluate the contribution of each metabolite to the axis of maximum variance of the data, here mainly representing the time evolution during the disease ( 1st Axis) . All results are always presented as mean±SD.

 For metabolites and the blood values one-way ANOVA (Prism 5.03, Graphpad, La Jolla CA USA) followed by Bonferroni's multiple comparisons post-tests were used (*$p < 0.05$, **$p < 0.01$, ***$p < 0.001$, ****$p < 0.0001$). Additionally, differences between brain regions were assessed using two-way ANOVA ($^{\$}p < 0.05$, $^{\$\$}p < 0.01$, $^{\$\$\$}p < 0.001$, $^{\$\$\$\$}p < 0.0001$). The % change was calculated in comparison to week-0. To test the correlations between the longitudinally acquired brain metabolites with brain Gln and blood NH4+ values (metabolites considered responsible for many of the changes in HE), Pearson correlation analysis was performed (Supplementary Material & methods).

For histology one-way ANOVA followed by post-hoc Turkey HSD was used to compare BDL and sham-operated rats (*$p < 0.05$, **$p < 0.01$, ***$p < 0.001$, ****$p < 0.0001$). All tests were 2-tailed.

## 2. Results

### BDL-induced differential brain regional changes in osmolytes, neurotransmitters and antioxidants measured by in-vivo $^1$H-MRS

Chronologically, the first common neurometabolic finding was an increase in Gln concentration at week-2 post-BDL. This was observed in the three brain regions, reaching significance at week-4 post-BDL (striatum +41%, hippocampus +63%, cerebellum +64%) with a rise in concentration observed until week-8 post-BDL. Gln increase was strongest in the ***cerebellum*** (+134% at week-8; **Fig. 1** A&B, **Fig. S4**). Brain Gln correlated strongly for all brain regions with blood $NH_4^+$ (p≤0.0001, **Fig. S5**).

As Gln increased the other main CNS osmolytes - Ins, Tau, creatine (Cr) and total choline (tCho=GPC+PCho) - decreased (**Fig. 1B**, **Fig. S4**). Specifically, **Ins** decreased consistently over time in the three regions reaching significance at week-8 post-BDL for ***cerebellum*** (-22%), ***hippocampus and striatum*** (-15%). **tCho** also decreased consistently in the three brain regions, with ***striatum*** showing the strongest reduction (-40%) at week-6 post-BDL continuing to decrease until week-8 post-BDL. **Tau** decreased consistently over time reaching significance at week-4 post-BDL for all brain regions (striatum -9%, hippocampus -8%, cerebellum -19%) and continued to decrease till week-8 post-BDL. This decrease was statistically strongest in the ***cerebellum*** when compared to the hippocampus and striatum. **Cr** also decreased consistently in all studied brain regions, most notably in the ***cerebellum*** (-9% at week-8 post-BDL). Ins and tCho correlated significantly with both Gln and blood $NH_4^+$ (in all brain regions, **Fig. S5**). Brain Gln showed a significant correlation with Tau for hippocampus and cerebellum while blood $NH_4^+$ correlated significantly with Tau in all brain regions. Similarly, Gln correlated significantly with Cr in hippocampus and cerebellum, while the correlation with blood $NH_4^+$ was not confirmed here.

Despite the smallest absolute increase in Gln, the ***striatum*** showed a more pronounced decrease of the absolute concentration of the main osmolytes (Ins+tCho+Cr+PCr+Tau) (8-weeks post-BDL: Gln +2.3 mmol/kg$_{ww}$ *vs* sum of osmolytes -4.2 mmol/kg$_{ww}$) than the hippocampus (8-weeks post-BDL: Gln +3.2 mmol/kg$_{ww}$ *vs* sum of osmolytes -2.6 mmol/kg$_{ww}$) and the cerebellum (8-weeks post-BDL: Gln +5.3 mmol/kg$_{ww}$ *vs* sum of osmolytes -4.1 mmol/kg$_{ww}$) (**Fig. 1B**).

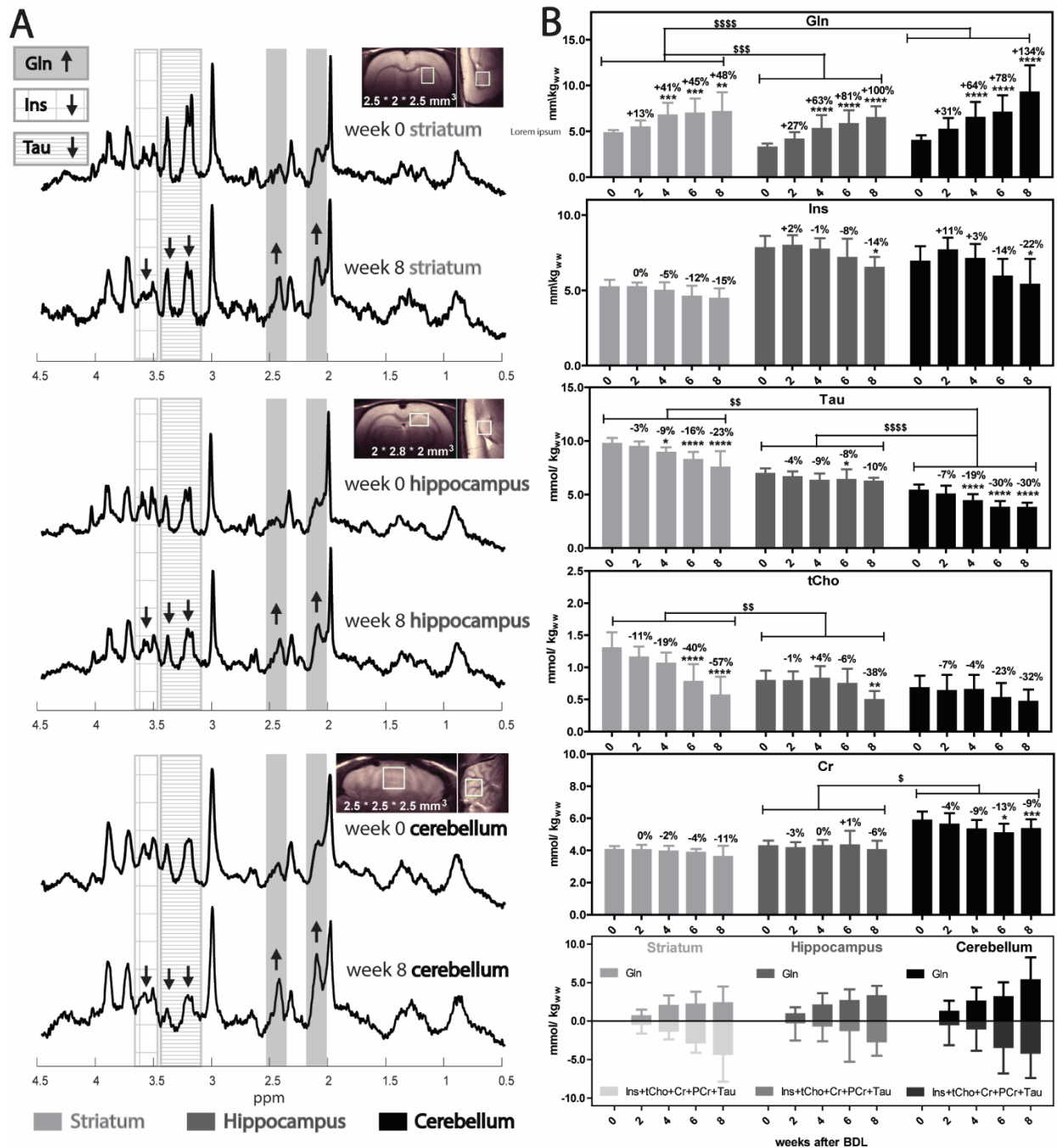

**Fig.1. (A) Representative *in-vivo* [1]H-MRS spectra** from 1 animal at week-0 (before BDL) and at post-operative week-8 in striatum, hippocampus and cerebellum. Increased Glutamine (Gln) and decreased Inositol (Ins) and taurine (Tau) can be observed. **(B) Longitudinal [1]H-MRS evolution of CNS osmolytes**. Last panel: Absolute Gln increase compared with the decreased sum of other

main osmolytes. Mean$\pm$SD; *p<0.05, **p<0.01, ***p<0.001, ****p<0.0001 (1-way ANOVA). Brain regions differences: $^\$$p<0.05, $^{\$\$}$p<0.01, $^{\$\$\$}$p<0.001, $^{\$\$\$\$}$p<0.0001 (two-way ANOVA)).

The neurotransmitter **Glutamate** (Glu) showed a decrease in the later stages of the disease, reaching a significant -10% at week-6 post-BDL in the ***striatum*** and continued to decrease until week-8 post-BDL in the three brain regions (**Fig. 2**, **Fig. S6**). Glu correlated stronger with blood $NH_4^+$ (significant for the three brain regions), than with Gln where the correlation was weaker showing significance for striatum and cerebellum only (**Fig. S7**). **Aspartate** (Asp, data not shown) and ***γ*-aminobutyric acid** (GABA) displayed no significant changes. However, GABA displayed an overall tendency of decrease especially in the hippocampus and cerebellum, and correlated significantly only with cerebellar Gln.

No significant changes in neuronal marker **NAA** were observed, however a tendency to decrease in the striatum was noted. **Lactate** (Lac) showed a striking increase of 84% only in the ***cerebellum*** in the final stage of the disease (week-8 post-BDL) (**Fig. 2**).

The antioxidant **ascorbate (Asc)** decreased significantly in the ***cerebellum*** at week-6 post-BDL (-32%). **Glutathione (GSH)** displayed no significant changes throughout the disease, with a tendency of increase in the ***cerebellum*** only (**Fig. 2**, **Fig. S6**).



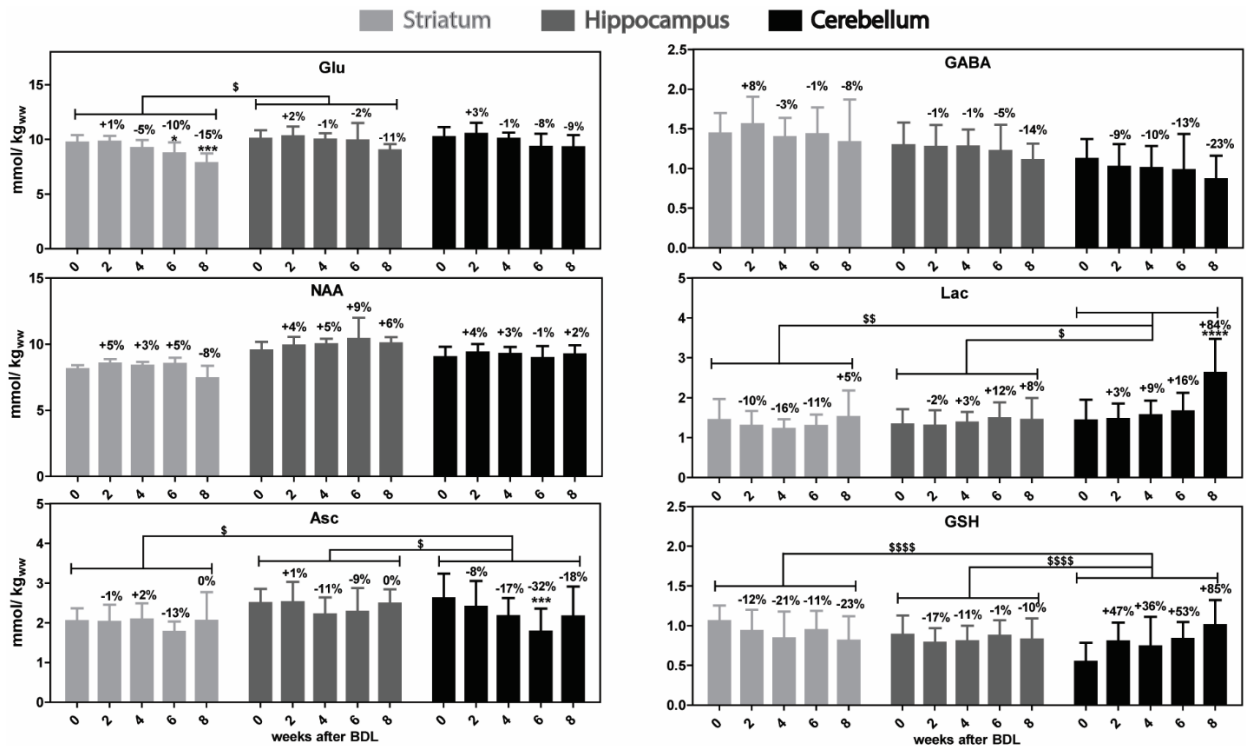

**Fig. 2. Longitudinal ¹H-MRS evolution of other brain metabolites**. Mean±SD; *p<0.05, **p<0.01, ***p<0.001, ****p<0.0001 (1-way ANOVA). Brain regions differences: $p<0.05, $$p<0.01, $$$p<0.001, $$$$p<0.0001 (two-way ANOVA)).

## Brain Gln and blood NH₄⁺ contribute to the variance in STATIS analysis of ¹H-MRS and blood data

The longitudinal measurements of metabolic profiles and blood biochemistry of all animals were analyzed altogether using STATIS (separated by brain region). The 1st Axis of principal component analysis represents the variance of the dataset brought on by the time evolution of the metabolites during disease. The metabolites with a high loading (positioned towards the extreme of the axis) are important, highly contributing to the 1st Axis and driving the variance. Increase of bilirubin and ammonia as well as decrease of blood Glc are known, longitudinally observed changes during CLD and confirmed here (**Fig. S3**) with their high contributions to the $1^{st}$ Axis (**Fig. 3**). Therefore, we can take them as a reference putting the contributions of brain metabolites into context of known changes during CLD. Gln has a high contribution to the variance (high contribution to the $1^{st}$ Axis) for all brain regions and behaves similarly to bilirubin and $NH_4^+$ (**Fig. 3**). On the opposite side of Gln we have Ins, Tau and tCho (GPC+PCho), which decreased as an osmotic response. The strong contribution of tCho to the $1^{st}$ Axis in striatum corresponds to its strongest decrease as shown in **Fig. 1B**. For the ***cerebellum***, a contribution of Lac is observed (on the same side as Gln) corresponding to a strong increase of Lac at week-8 in this brain region.

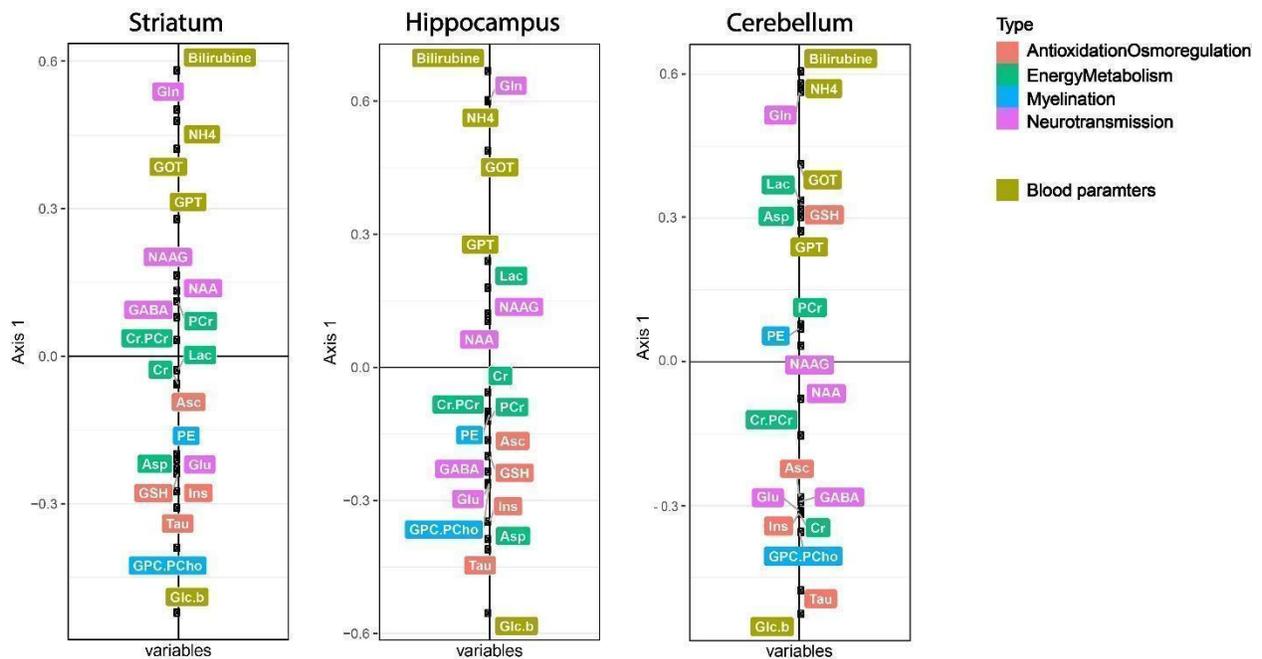

**Fig. 3. The first axis resulting from the STATIS analysis**. The axis corresponding to the dataset containing brain metabolites + blood parameters. Metabolites that highly contribute to the 1st Axis (towards the extremities) are the ones driving the variance and they can be grouped based on their contribution. The variables that are positioned closer to the extremes of the axis are the ones highly

contributing to the variance explained by the axis (variance of the longitudinal evolution of metabolites during the disease in a specific brain region), while the variables positioned at the opposite extremities showed opposite changes.

BDL-induced brain regional changes in astrocytes, microglia and neurons

A significant increase in astrocytes (GFAP+ cells, astrocytosis-reactive astrocytic response) and total (all) cell nuclei number was observed at 4-weeks post-BDL in the **hippocampus** (+48%, 45%), **cerebellum** (+45%, 53%) and **striatum** (+31%, 21%), respectively (**Fig. 4B**). Although a significant reduction of astrocytosis was observed at week-8 when compared to week-4 post-BDL (hippocampus: +48% to + 8%, cerebellum: +45% to +23%, and striatum: +31 to +15%, **Fig. 4B**), the number of astrocytes remained slightly increased compared to sham. Nuclear staining showed the presence of vesicular nuclei with migrated chromatin, typical for the Alzheimer type II astrocytes at 4-weeks post-BDL surgery in all brain regions (**Fig. S8**). Sholl analysis[13] depicted a time-dependent reduction of the number of processes (GFAP+ astrocytic processes) at week-4 (hippocampus -6%, cerebellum -32%) and continued to decrease until week-8 post-BDL (**Fig. 5B**) with cerebellum showing a stronger reduction in number of processes. In striatum the decrease became significant only at week-8 (-12%). The length of GFAP+ astrocytic processes decreased over time reaching significance at week-4 for all brain regions (hippocampus -13%, cerebellum -17%, striatum -18%, **Fig. 5B**). A significant decrease in the GFAP+ astrocytic processes intersections in the hippocampus and cerebellum was observed at 4-weeks, displaying a stronger decrease at week-8 (hippocampus: -14%-Ring1, -39%-Ring2, -72%-Ring3, cerebellum: -18%-Ring1, -27%-Ring2, -59%-Ring3, **Fig. 5C**). No significant decrease of intersections number was observed for striatum (**Table S4**, **Fig. S9**). Bergmann glia (BG) processes exhibited also morphological alterations. Bergman glia (BG) processes displayed a significant increase in appendages at 4-weeks post BDL (**Fig. 4C**).

Microglia and macrophage activation were depicted using Iba1 staining. A significant increase in Iba1+ cells was detected at 4-weeks post-BDL (hippocampus +39.9%, cerebellum +27.2%, striatum +38.6%, (**Fig. 6A&B**, **Fig**. **S10**). At week-4 post-BDL, microglia were morphologically altered, with increased process length and cell perimeter (hippocampus: +18.6%, +10%, cerebellum: +27%, +18.5%, striatum: +16%, +9.8%, processes length and cell perimeter, respectively) (**Fig. 6C**). At week-8 post-BDL a significant reduction of microglia Iba1+ cells was observed in all brain regions and was accompanied by a significant increase of amoeboid microglia number (**Table S4**).

In sham, sparse Iba1+ phagocytically active macrophages were located in the choroidal stroma, in the 3$^{rd}$ and lateral ventricle and hindbrain (aqueduct) choroid plexus. Starting 4-weeks post-BDL, a significant increase in Iba1+/activated macrophages was identified in the stroma as well as in the base of choroidal epithelial cells, with an increase in activated macrophages in the lateral ventricle at week-8 post-BDL compared to week-4. At week-8 post-BDL clots like macrophages formation were present, in the 3$^{rd}$ and lateral ventricle, and hindbrain choroid plexus (**Fig. 7**, **Fig. S10**).

Changes in neurons structure was revealed using Golgi-Cox staining: a significant increase in CA1 (+67%) and DG (+54%) neuronal soma surface and a significant loss of dendritic spines density in CA1 (-49%, apical and basal) and DG (apical - 43%) hippocampal neurons (**Fig. 8A, Fig. S11, Table S5**). In contrast to hippocampal neurons, Purkinje cells in cerebellum showed a significant decrease of the neuronal soma surface (-22%) and dendritic spines density depletion (apical -24% **Fig. 8B**).

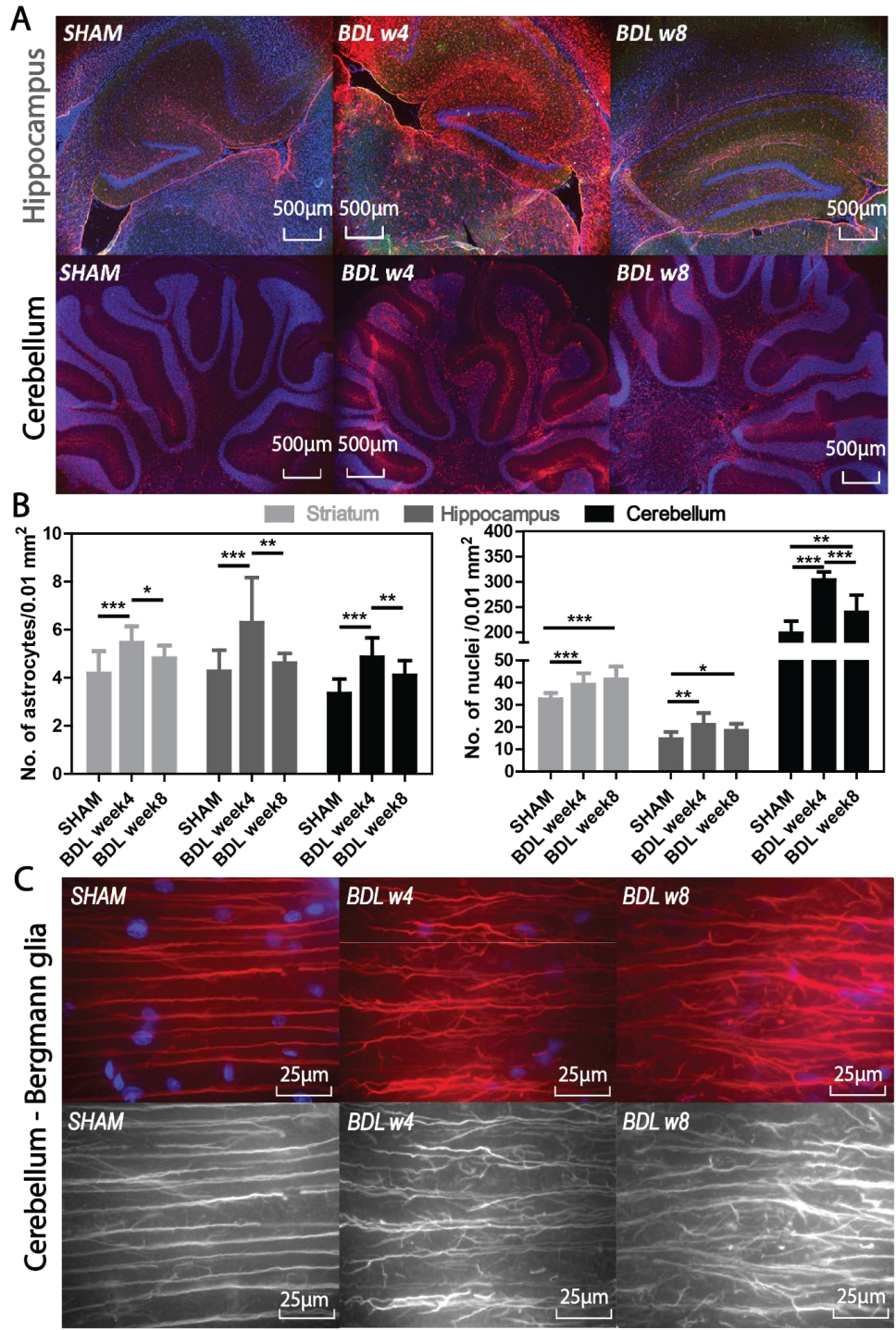

**Fig. 4. Anti-GFAP (red) and DAPI (blue) staining of hippocampus and cerebellum: (A)** Representative micrographs of sham, week-4 and 8 post-BDL, and **(B)** the astrocytes density quantification. Note the significant increase in astrocytes number at week-4 post-BDL. **(C)**

Bergmann glia: fibers deterioration and expansion of end-feet through the pial surface starting at BDL week-4 that increased in severity at week-8. One-way Anova with post-hoc Tukey HSD, *p<0.05, **p<0.01, ***p<0.001, ****p<0.0001, mean ± SD.

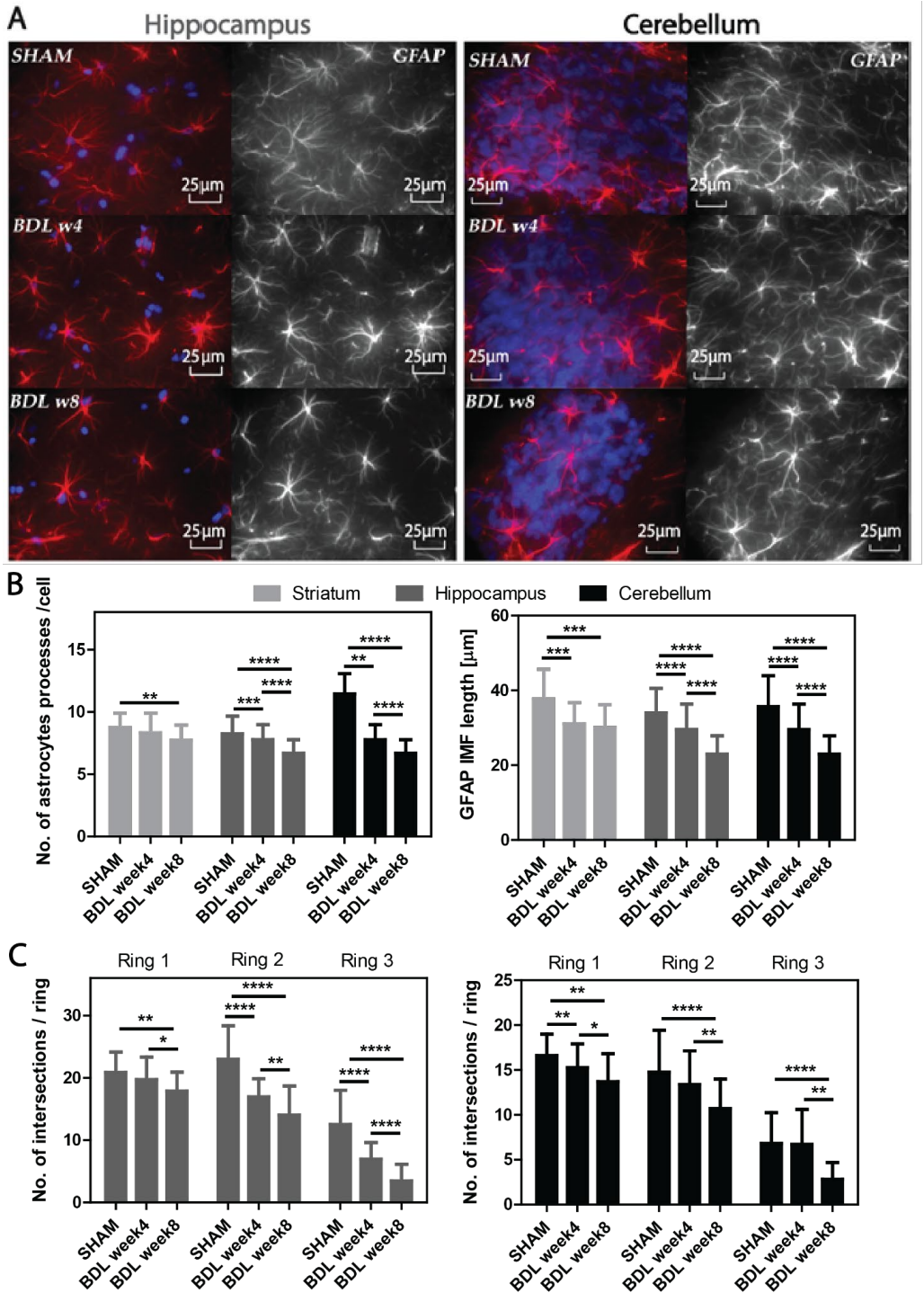

**Fig. 5: Anti-GFAP (red) and DAPI (blue) staining. Astrocytic morphology analysis: (A)** Brain sections from sham and BDL rats at 4 and 8-weeks post-BDL (hippocampus and cerebellum). **(B)** Astrocytic processes number (left panel) and process lengths (right panel) quantification, and **(C)**

the number of intersections across the given rings starting from the cell body's center - plotted according to subregions (1-3) (Sholl analysis, **Fig. S1**). One-way Anova with post-hoc Tukey HSD, *p<0.05, **p<0.01, ***p<0.001, ****p<0.0001, mean ± SD.

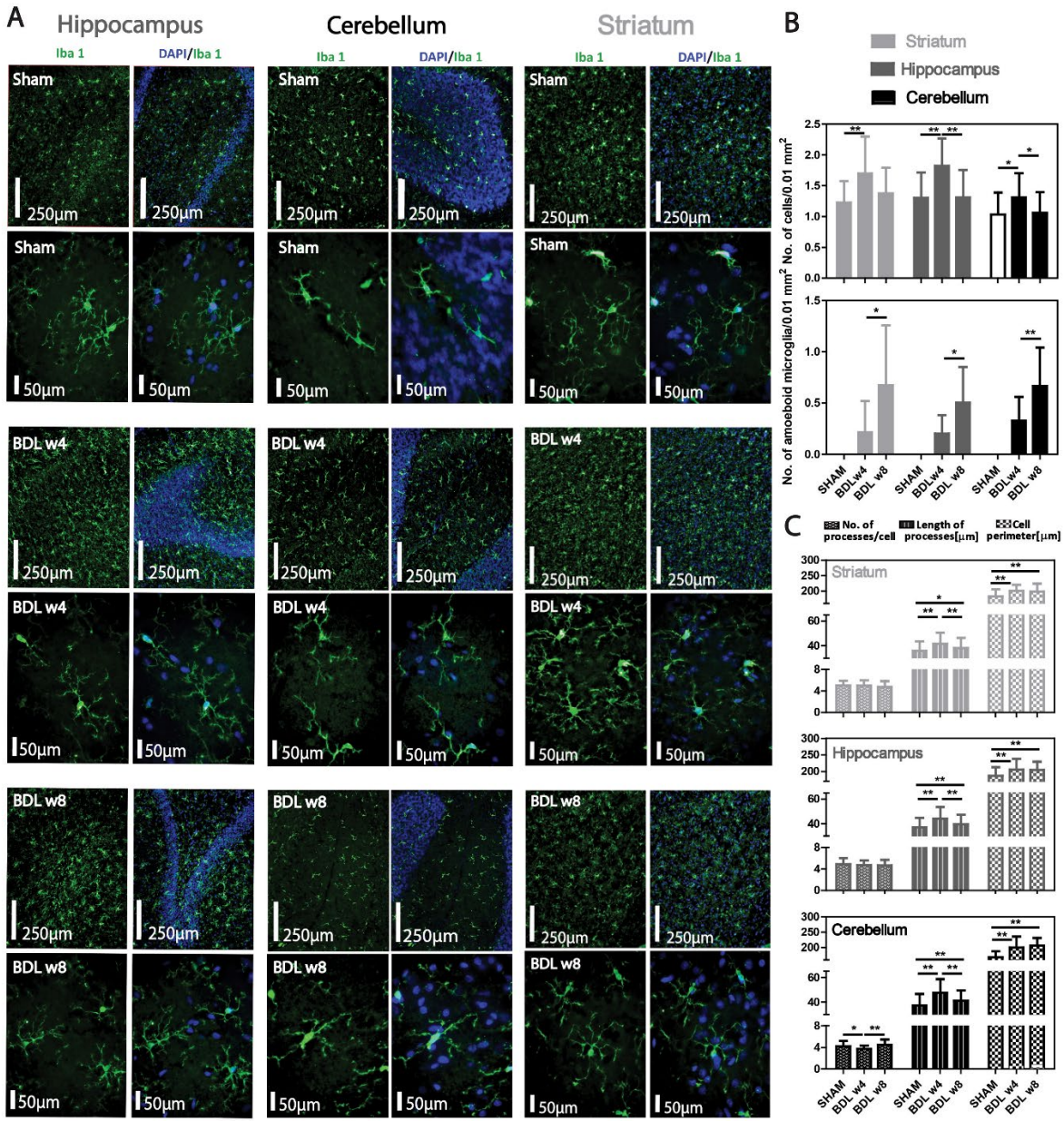

**Fig. 6. Anti-Iba1 (green) and DAPI (blue) staining of microglia: (A)** Brain sections from sham and BDL rats at 4 and 8-weeks post-BDL (hippocampus, cerebellum and striatum). **(B)** Microglia

density and amoeboid microglia quantification. **(C)** Microglia processes number, processes length and the cell perimeter analysis. One-way Anova with post-hoc Tukey HSD, *p<0.05, **p<0.01, ***p<0.001, ****p<0.0001, mean ± SD.

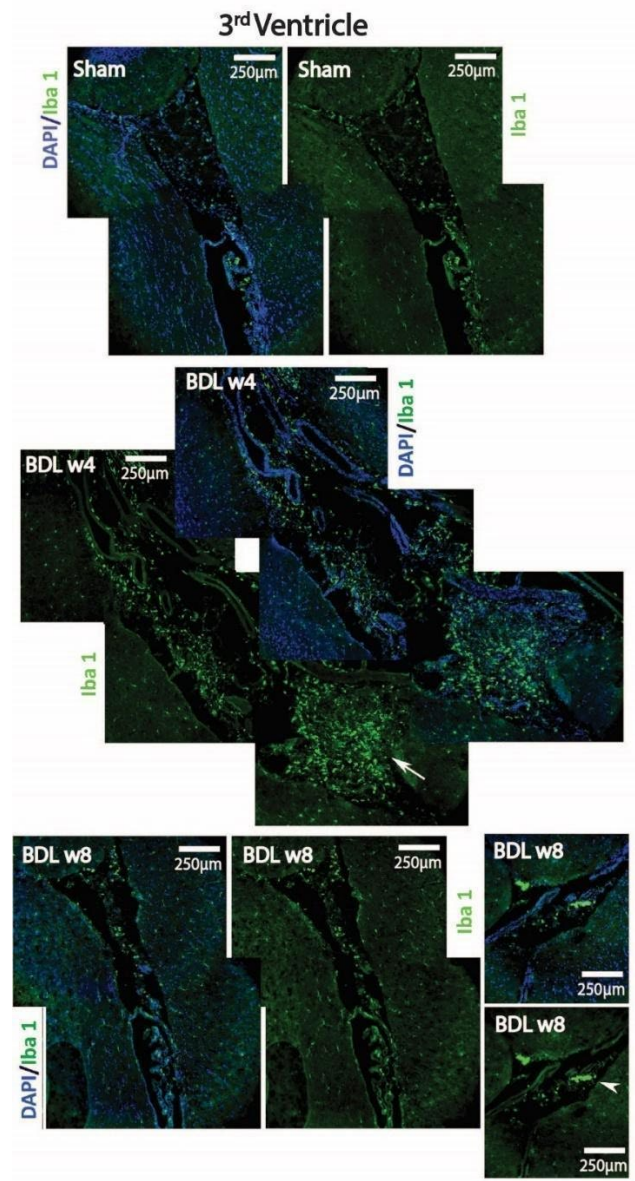

**Fig. 7. 2D reconstruction of anti-Iba1 (green) and DAPI (blue) staining of diencephalon (3<sup>rd</sup> ventricle):** Activated macrophages accumulation and cloth formation. Arrow - increased number of macrophages, arrowhead – macrophages clots formation.

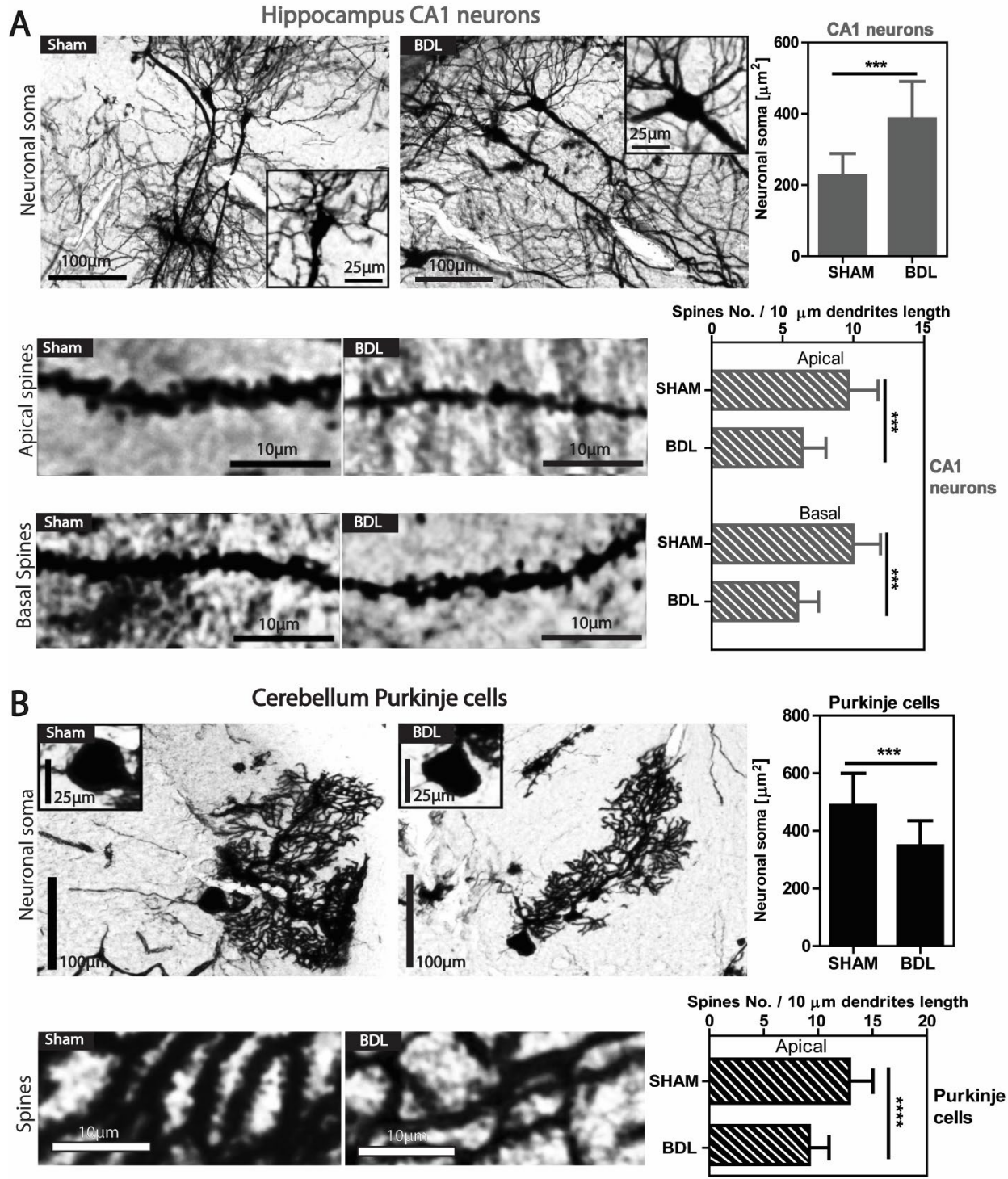

**Fig. 8. Golgi-Cox staining:** Representative micrographs of Golgi-Cox staining: neuronal morphology analysis. In BDL the hippocampal pyramidal CA1 neurons **(A)** and cerebellum

Purkinje cells **(B)** showed altered soma size and decreased dendritic spine density. One-way Anova with post-hoc Tukey HSD): *p<0.05, **p<0.01, ***p<0.001, ****p<0.0001.

3. Discussion

In this unique study combining longitudinal *in-vivo* [1]H-MRS in a recognized rat model of CLD induced type C HE and *ex-vivo* analysis of brain cell subtypes, we showed that there are shared and also differential metabolic and cellular responses to CLD in the cerebellum, hippocampus, and striatum. **Metabolically** Gln displayed the most rapid and stronger change in all the brain regions, thus was pointed as a surrogate of the disease. This was followed by a decrease in osmolytes, neurotransmitters and changes in antioxidants across regions. Moreover, cerebellum was the region with the heaviest burden of Gln, with some distinct changes such as increase in Lac and GSH, and decrease in Asc and GABA, while the striatum displayed the lowest increase in brain Gln but the strongest decrease in brain osmolytes (e.g. tCho). **At the cellular level**, apart from the well-known astrocytic alterations characteristic of HE, several regional differences were observed. In the cerebellum, the number of processes was strongly reduced compared to other regions. Furthermore, we observed morphological alterations in microglia, including changes in processes length and perimeter, and in neurons, with reduced soma size of Pj cells (cerebellum), increased soma size of CA1 and DG (hippocampus), and decreased spine density (slightly stronger in the hippocampus). Moreover, the striatum seemed to display the weaker changes in astrocytes morphology potentially due to lower Gln increase but stronger osmotic answer, something requiring further investigations.

*Common changes among the three studied regions*

The longitudinal measurements of CNS metabolic profiles and blood biochemistry helped identify the brain regional and blood 'drivers' of the disease. The longitudinal measurements of blood showed a rapid systemic rise of bilirubin and $NH_4^+$, as previously observed in CLD[13,16]. In addition, longitudinal [1]H-MRS measurements in three separate brain regions (striatum, hippocampus, and cerebellum) allowed us to uncover global, common neurometabolic alterations. We could therefore identify Gln as the first observed globally increased metabolite in all three brain regions. Moreover, the STATIS analysis revealed that Gln, $NH_4^+$ and bilirubin, similarly drive the variance of the dataset (**Fig. 3**) confirming that $NH_4^+$ increase in the blood induces a very rapid cascade of metabolic events in the CNS initiated by Gln increase. Moreover, a recent study on treated BDL rats (rifaximin and the probiotic VIVOMIXX[17,18]) identified Gln as the fastest responding metabolite to treatment (lower increase), even when no $NH_4^+$ response was detected. This highlights the importance of Gln in type C HE and the value of *in-vivo* methods detecting it to study the effect of therapies[19]. In the same context, we recently demonstrated that children with chronic compensated liver disease do not show neurometabolic changes measured by [1]H-MRS at 7T[20], while a significant increase in glutamine was observed in a child with type C HE (extra-hepatic congenital portosystemic shunt) compared to controls[21].

The brain is very susceptible to excess ammonia and it primarily depends on its clearance by Gln synthesis (via glutamine synthetase enzyme) in astrocytes to prevent secondary neurological damage[22]. In disease condition the observed elevated Gln is largely situated in astrocytes causing an osmotic imbalance and a gradual decrease of other brain osmolytes[13,16,23] (i.e. Ins, tCho, Tau and Cr **Fig. 1B**). Ins is a molecule found predominantly in glial cells where it functions as an osmolyte[13], decreasing in response to increased Gln levels. tCho, is also considered a glial marker, as evidenced by variations in its diffusivity linked with microglial reactivity[24], which supports its decrease in response to Gln rise. Tau and Cr can, among other functions, act also as osmolytes, especially in the presence of strong Gln osmotic pressure[13,25]. Taken together, this neurometabolic pattern was observed in all of the investigated brain regions and can be considered another disease-wide feature.

The observed increase in number of astrocytes at 4-weeks post-BDL was one of the first global (in all studied regions) cellular changes. This increase is a hallmark of astrocyte reactivity and could be related to the stimulation of mature astrocytes and reentering into the proliferation cycle, a common feature in neuropathological disorders[26]. In parallel, the osmotic stress together with OS and other factors, likely contributed to the reported astrocytic morphological changes (processes length and number)[27]. It has been shown that GFAP expression decreases with HE progression[28], as also shown here at week-8 post-BDL, and that the altered astrocytes morphology disrupts their function[13,29]. Astrocyte morphological changes could trigger astrocytic and then brain functional deficits[13,29].

Microglia activation is associated with a rise in pro-inflammatory cytokines (i.e.,TNF, IL-1β, IL-6) and release of ROS[30]. This cellular alteration was observed in all investigated brain regions. Furthermore, we detected the appearance and rise in ameboid microglia numbers consistent with our prior observation of brain IL-6 accumulation[3]. The presence of amoeboid microglia has been associated with neurodegenerative disorders such as Alzheimer's, Parkinson's, and Huntington's disease, all of which are characterized by chronic neuroinflammation[31]. Their increased number may also be linked to an increase in their phagocytic activity (phagocytosis of cellular debris)[31]. These findings are indirect evidence of neuroinflammation and suggest a possibility of brain degeneration in type C HE[32]. In addition, an important activation of choroidal macrophages was observed, which secrete cytokines and ROS, and infiltrate into the CNS[33]. Changes in glucose and Gln metabolism alter macrophage signaling, resulting in the formation of M1 (pro-inflammatory) macrophages, as reported herein, and release of pro-inflammatory cytokines (IL-6, IL-1, and TNF)[34]. Therefore, choroid plexus may also play an important role in modulation of neuroinflammation in type C HE.

The [1]H-MRS measurements additionally showed an overall decrease of Glu, an alteration that can have multiple causes, including increased synthesis of Gln from $NH_4^+$ which could contribute to altered neurotransmission[35], or an $NH_4^+$ induced OS which might inhibit Glu uptake in astrocytes[36]. Glu is a direct precursor of GABA synthesis via the enzyme glutamate decarboxylase[37,38]. We observed a decline of GABA concentration which was more pronounced in the cerebellum. Our finding is supported by previously[13] and here observed Glu decrease which would result in reduced GABA synthesis. GABA decrease in cerebellum could also be linked to the changes observed herein in the Purkinje cells. However, other studies have reported an increased GABA synthesis[39] or increased GABA concentrations in BDL rats and HE patients[40], respectively. These controversial findings together with the major involvement of GABA in brain cognitive function[41] make it an important target for future investigations.

The majority of HE research to date has centered on astrocyte morphology and biology, with minimal focus on neurons[11,42,43]. We reported pronounced changes of the hippocampal and cerebellar neuron morphology at week-8 post-BDL. Our data showed for the first time a significant and regionally differential change in soma size of hippocampal CA1 and DG neurons, and the cerebellar Purkinje cells. A stronger decline of the spine density in the cerebellar Purkinje cells may relate to a significant increase of CNS OS[4], as highlighted by the changes in antioxidants, and to Glu decrease, which plays a significant role in the activity-dependent plasticity of astrocyte neuron interactions[35]. Only a few recent studies briefly explored neurons in the context of type C HE[11,43], their findings suggesting the presence of neurodegeneration and neuronal cell loss, in support of our findings. Given that neuronal morphological changes can be linked with memory decline and cognitive dysfunction[44], these results offer development of a novel research direction in HE.

*Differential brain regional changes*

*Cerebellum*

Although Gln increased globally in the brains of BDL rats, the cerebellum showed the strongest increase, suggesting that cerebellar astrocytes had to bear the heaviest Gln load. In parallel, stronger astrocytic (astrocytic processes) and microglial changes combined with a decreased Pj cell soma and spine density (**Fig. 8B**) were observed, suggesting that this cellular response may be a distinct local response to the Gln load. Tau showed a significantly stronger decrease in the cerebellum than in the other brain regions and correlated with the Gln increase, suggesting that its role in astrocytic osmoregulation[45] is more prevalent in this brain region due to the heavier load of brain Gln. It could be argued that the high cerebral blood flow and ammonium extraction factor reported in cirrhotic patients[4] together with the comparatively high percentage of neurons and astrocytes in the cerebellum[46], make this brain region more susceptible to ammonium. Additionally, the cerebellum's tendency for stronger neuroinflammation[3,4,8,47] makes it increasingly affected by inflammatory cytokines, among other deleterious molecules[3].

Cr also showed a stronger decrease, probably assuming it's osmoregulatory role[48]. Additionally, the high exposure to $NH_4^+$ can cause an inhibition of Cr synthesis in the CNS, contributing to its decrease[49]. As such Cr decrease in BDL rats will potentially impact energy metabolism due to its important role in the regeneration of ATP through the Cr/phosphocreatine/creatine kinase (Cr/PCr/CK) system[50]. Moreover alterations in [1]H-MRS detectable tCr or Cr were associated with creatine transporter density[47,49], suggesting alterations in neuronal energy metabolism.

The behavior of Lac was significantly different in cerebellum with a peak at week-8 (+84%), a unique change confirmed by its higher contribution to the 1[st] axis of STATIS analysis (**Fig. 3**). The previously published results of Lac changes are contradictory: increase in Lac playing a role in pathogenesis of brain edema in BDL rats[51] or decrease of cortical extracellular Lac pointing towards its intracellular accumulation[52] and decrease in Lac cortical concentration[53]. However, a brain region specific lactate behavior should not be excluded, as measured in the current study. Increased lactate production in astrocytes that feed the surrounding neurons, as additional fuel to Glc, under glutamatergic synapse activation (astrocyte-neuron lactate shuttle hypothesis)[54], has been suggested as a possible pathway for Lac increase in HE[55]. In contrast, we observed a decreased density of dendritic spines of Purkinje neurons (forming glutamatergic synapses with parallel fibers from granular neurons) and a decreased Glu under BDL, suggesting a lower activity of the glutamatergic synapses. Moreover, in the present and previous studies we showed a highly decreased blood Glc[3,56] significantly driving the data variance (**Fig. 3**), with a twofold decrease in glucose uptake measured in the same animal model in cerebellum and hippocampus *in-vivo*[56]. Therefore, none of our findings support the increased lactate production by astrocytes to support neurons, as this would require increased Glc uptake. Altogether, our data most probably suggest that the significant increase of Lac in the cerebellum is the reflection of intense cellular stress, both neuronal and glial, at this late stage of disease, i.e. 8-weeks post-BDL. However, we cannot exclude that alternative substrates to Glc (such as Gln, ketone bodies or fatty acids) could be used in the TCA cycle earlier in disease development, warranting further investigations.

The first line antioxidants, Asc and GSH, maintain the redox homeostasis and keep the balance between ROS generation and elimination[3]. A significant decrease in Asc concentration and a unique increase in GSH were observed in cerebellum (**Fig. 2**), highlighting it again as the region with stronger neurometabolic changes. The UHF $^1$H-MRS gives us an exclusive opportunity to measure *in-vivo* Asc (highly concentrated in neurons[57]) and GSH (high concentration in astrocytes[27]), the neuroprotective antioxidants which provide a quick response to OS[3,57]. The astrocyte activation as shown herein is an important element of brain immune response which may significantly compromise the antioxidant capacity[27]. Our data showed brain redox homeostasis alterations in the cerebellum and corroborate our previous findings of direct *ex-vivo* detection of increased OS and significantly increased GPX-1 in the cerebellum of the BDL rat[3], and previous post-mortem studies in patients[42].

*Striatum*

The pattern of Gln increase differed in striatum from the other studied regions, together with smaller astrocytic alterations. While in cerebellum and hippocampus Gln was consistently increasing over time, the Gln concentration increase started to decelerate at week-4 in the striatum. Interestingly the main CNS osmolytes continued to strongly decrease after week-4 post-BDL. Regardless of the lower Gln-induced osmotic pressure (than in cerebellum). Tau showed a strong and significant decrease starting from week-4 post-BDL (**Fig. 1B**). This behavior can be explained as a delayed osmotic response or suggest that Tau assumes a different or an additional role (antioxidant/anti-inflammatory) to the osmoregulatory one[45] (supported by a significant correlation of Tau with $NH_4^+$ but not with Gln, **Fig. S5**). It was shown previously that Tau plays a role in striatal plasticity, exerting a neuroprotective role under metabolic stress[58]. Therefore, a strong change in Tau may have molecular implications in this brain region. tCho showed the strongest and significantly different decrease in striatum compared to the other brain regions. Apart from its osmoregulatory role tCho is required for membrane phospholipid synthesis[59] and myelination, therefore its strong decrease could be considered as a sign of neurodegeneration.

4.  Conclusion

In an established model of type C HE, both common and differential metabolic and cellular changes were identified across brain regions. Common metabolic changes included elevation of Gln-chronologically first-, decreases in osmolytes, and a decrease in neurotransmitters to varying degrees. At the cellular level, glial changes included an increase of astrocytes number, a decrease in number and length of astrocytic processes while microglia displayed increased cell perimeter and processes length, and an overall rise in number of microglia and amoeboid cells. Importantly, neurons showed a decreased spine density and soma size change (decreased soma size of Pj cells in cerebellum vs increased CA1 and DG neurons in hippocampus). Further, each anatomical structure studied was characterized by its own metabolic and cellular signature. The cerebellum displayed the highest Gln burden, together with significant elevation of Lac and decreased GABA, as well as changes in antioxidants concentration. This was paralleled by more pronounced astrocytic changes and decrease soma size of Pj cells, arguably related to the osmotic stress driven by Gln load. Next, the striatum displayed the lowest increase in brain Gln but the strongest decrease in brain osmolytes, potentially underlying the milder changes in astrocyte morphology. Taken together, these results suggest that although there are some common metabolic and cellular responses in type C HE, brain regions show a differential susceptibility, possibly related to cellular composition, although the teleological significance is unclear.

Based on the early rise of Gln in all brain regions studied, it is tempting to hypothesize that Gln is the molecular contributor to a cascade of cellular changes though osmotic, oxidative stress and neuroinflammation. Moreover, the observed neuronal alterations suggest an interdependence between astrocytic and neuronal morphological changes which may affect the entire neuronal network activity during disease progression. The specific impact of these events on neurodegeneration and functional deficits observed in HE still needs to be clarified.

# Differential metabolic and cellular brain regional vulnerability in adult rats with chronic liver disease

Dunja Simicic, Katarzyna Pierzchala, Olivier Braissant, Dario Sessa, Valérie A. McLin, Cristina Cudalbu

**Table of contents**

**Supplementary Materials and Methods**

*BDL rat model of CLD induced-type C HE*

All experiments were approved by the Committee on Animal Experimentation for the Canton de Vaud, Switzerland (VD3022/VD2439), were conform to the Animal Research: Reporting of In Vivo Experiments (ARRIVE) guidelines (http://www.nc3rs.org.uk/arrive-guidelines) and all animals received human care. Wistar male adult rats (n=45, 175-200g, Charles River Laboratories, L'Arbresle, France) were used: 35 rats underwent BDL surgery, a model of type C HE accepted by ISHEN[1], while the rest was Sham operated (BDL surgery protocol: https://doi.org/10.5281/zenodo.10652104). Animals were housed in cages (2-3 rats per cage) in the animal facility at 20–24° C with a 12 h light/dark cycle, and free access to water and food (SAFE 150 SP-25, SAFE, U8409G10R). Information about the number of animals for each experiment is provided in **Tables S1-S3**.

*Biochemical measurements to validate the CLD*

Liver parameters (plasma bilirubin, aspartate aminotransferase (AST/GOT) and alanine aminotransferase (ALT/GPT) (Reflotron Plus system, F. Hoffmann-La Roche Ltd.)), glucose (Glc) (CONTOUR®XT) and blood $NH_4^+$ (PockketChem[TM] BA PA-4140) were measured longitudinally every 2 weeks (week 0, 2, 4, 6, 8) (**Table S1**).

*In-vivo metabolic changes using [1]H-MRS*

[1]H-MRS spectra were acquired on the 9.4 T system (Varian/Magnex Scientific) using the SPECIAL sequence (TE=2.8ms) as previously described[11]. Three different volumes of interest (VOI) were selected in hippocampus ($2x2.8x2mm^3$), cerebellum ($2.5x2.5x2.5mm^3$) and striatum ($2.5x2x2.5mm^3$). LCModel was used for quantification with water as internal reference allowing the quantification of a total of 18 metabolites. The scans were performed before (week 0) and after BDL at weeks-2, 4, 6, and 8, thus each animal was its own control. For [1]H-MRS from week-0 to week-6, 8 to 23 rats per week were scanned while at week-8 the number of scanned rats varied between the regions (cerebellum n=8,

hippocampus n=7, and striatum n=4) (**Table S2**). Animals were sacrificed at different time points for histology measurements or because they reached the end point allowed by the Committee on Animal Experimentation for Canton of Vaud (e.g. 15% of weight loss), justifying the different number of rats at each time point.

*Cellular morphological changes using histology*

**Tissue Preparation:** Animals were deeply anesthetized with 4% isoflurane for 5 min and an analgesic was injected (Temgesic (ESSEX) (0.1 mg/kg)) before transcardial perfusion with cell culture medium RPMI 1640 (pH 7.4, Sigma), supplemented with 10% of Fetal Bovine Serum (FBS, Sigma) and 1% of antibiotics (50.5 units/ml penicillin, 50.5 μg/ml streptomycin and 101 μg/ml neomycin, Sigma) to wash out blood and keep the brain cells alive, pH 7.4.

**Immunohistochemistry (IHC):** Animals were sacrificed for histological evaluation at week-4 (BDL, n=3) and 8 (BDL, n=3 and Sham, n=3) post-surgery (**Table S3**). Brains were removed and fixed in 4% formaldehyde PBS solution overnight at 4°C, followed by PBS wash and 48h cryopreservation in 30% sucrose in PBS at 4°C. They were embedded in an optimum cutting temperature compound (Tissue-Tek® O.C.T. Compound) then flash-frozen using dry ice and 2-Methylbutane (Sigma) for sectioning. Sections were mounted on Superfrost Plus microscope slides (Thermo Scientific). After the staining procedure, the slides were mounted with ProLong™ Diamond Antifade Mountant (P36970, Thermo Fisher) and covered with coverslip.

Mouse monoclonal anti-GFAP antibody (MAB360 Merck Millipore) (2 h at RT) at 1/100 dilution with secondary Alexa Fluor® 594-AffiniPure Rat Anti-Mouse IgG (H+L) antibody (415-585-166 Jackson ImmunoResearch Europe Ltd.) (1 h at RT) at 1 / 200 dilution were used to characterize the astrocytes morphology[11]. Iba1 goat polyclonal antibody (PA5-18039 Thermo Fisher) (2 h at RT) at 1/100 dilution with Alexa Fluor 555 - Rabbit Anti-Goat IgG (H+L) Superclonal™ secondary antibody, (A27017,

Thermo Fisher) (1 h at RT) at 1 / 2500 dilution were used to characterize the microglia morphology. Nuclei were stained with DAPI (D1306, Thermo Fisher).

**Fluorescence microscopy** examination of tissue sections was performed. Rats brains (week-4 post-BDL (n=3), week-8 post-BDL (n=3) and Sham (n=3), **Table S3**) were sliced at 16 μm thick sagittal sections using Leica Jung CM 1800 cryostat. For each rat, seven different slides were analyzed (distance between tissue sections ~250μm), resulting in a total of 63 slides prepared. To investigate the morphometrical changes of astrocytes and microglia in the hippocampus, cerebellum, striatum, the Sholl analysis was implemented (**Fig. S1**)[2]. For the astrocytes each process was separated by segment order – concentric spheres originating from the soma (**Fig. S1A**). A total number of ~200 representative astrocytes of each group were randomly traced for all processes revealed by GFAP staining. An average of 350 astrocytic processes per each sample were measured (~1000 processes per each group). The number of processes leaving the cell body, total length and intersections of processes were quantified and plotted according to subregions (1-3) (**Fig. S1A**). The microglia (**Fig. S1B**) cell body perimeter was determined by encircling the microglial using the polygon tool to connect distal extremities of every process. A total number of ~50 representative microglia of each group were randomly traced for all processes revealed by Iba1 staining. An average of 100 microglial processes per each sample were measured (~300 processes per each group). The number of primary processes and total length of all processes were measured.

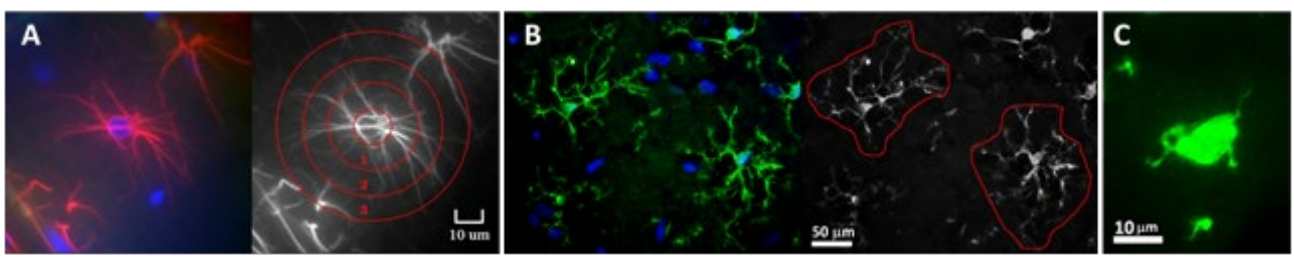

**Fig. S1. Sholl analysis: (A)** Astrocytes - morphological characterization of the number of intersections of branches with radii at various distances from the cell body. **(B)** Microglia - morphological characterization of the cell perimeter, and **(C)** Amoeboid microglia are distinguished as big, spherical stained cells with few, short or no processes.

**Golgi Cox staining[3,4]** was performed to unveil the detailed morphology of the hippocampus CA1 neurons and DG granule cells (**Fig. S2**), and cerebellar Pj cells. Brains (week-8 BDL (n=7) and Sham (n=6), **Table S3**) were cut into 115 µm thick sagittal sections using Leica VT1200 S vibratome. Tissues with homogeneous staining and clearly visible dendritic segments and spines were chosen for quantitative analysis (25 slides/hemisphere). The surface of the neuronal soma was measured, and manual counting of the dendrite spine was conducted[5]. Measurements were obtained from the CA1 neurons (BDL soma ~200 cells, apical and basal dendrites ~100 each; Sham soma ~120 cells, apical and basal dendrites ~60 each) and DG granule cells (BDL soma ~200 cells, apical dendrites ~150; Sham soma ~180 cells, apical dendrites ~80) of hippocampus, and Pj cells (BDL soma ~500 cells, apical dendrites ~220; Sham soma ~180 cells, apical dendrites ~80) in cerebellum. Sections were digitized using a MEIJI TECHNO TC5600 microscope with an INFINITYX-32 camera. Images were acquired and analyzed using the INFINITY ANALYZE 7 software (Lumenera, Canada).

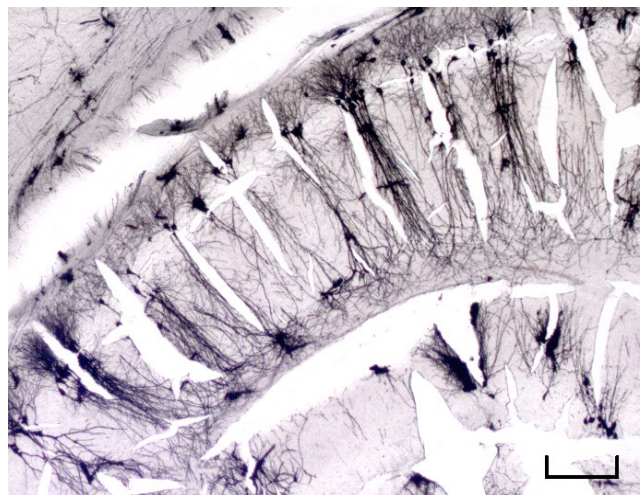

**Fig. S2. Representative micrograph of Golgi–Cox impregnated hippocampus** (scale bar = 200 µm).

*Statistical analysis*

**[1]H-MRS** data are displayed as scatter plots showing the relative increase of metabolite concentrations over time during disease progression. The plotted values were obtained by subtracting the week 0 concentration of a metabolite for every rat and brain region from the corresponding week 2, 4, 6 and

8 measured concentrations, and therefore keeping only the relative (to week 0) change in the disease. In this way the starting differences in metabolite concentrations between the brain regions[6,7] were removed, describing only the metabolite evolution due to the disease while clearly pointing to the brain regional difference for the concerned metabolites. The plots were created in RStudio[8] where additionally a LOWESS (Locally Weighted Scatterplot Smoothing) regression was applied and plotted to better visualize and distinguish data trends between brain regions.

**STATIS:** The pipeline for application of the STATIS method was applied as previously described[9,10] using RStudio[8]. STATIS method is created as an extension to PCA, designed to manage multiple data tables that contain set of variables collected on the same observations[11,12].

As described the animals were scanned (in three brain regions) and the blood sampling was performed longitudinally at week 0, 2, 4, 6 and 8. For one brain region this provides us with several quantified metabolites for each rat at five different time points. Therefore, we have a three-dimensional dataset suitable for the STATIS analysis. The analysis was performed on each brain region individually on a combined dataset consisting of the brain metabolites and blood parameters.

The data was prepared by stacking rats at different time points as rows and metabolites as columns (for each brain region). This resulted in three datasets: brain metabolites + blood parameters – striatum, brain metabolites + blood parameters – hippocampus, brain metabolites + blood parameters – cerebellum. Every dataset was processed using the following procedure. The matrix/dataset (rats at different time points -rows, metabolites -columns) was centered by subtracting the mean of every column to the corresponding column (e.g. the mean value of Gln measured for all the rats at all time points was subtracted from every Gln measurement). The few missing values due to low spectral quality or missing a scan were imputed using missMDA package (RStudio).

**Table S1. Detailed report on each rat used for biochemical measures.**

| | Blood ($NH_4^+$) - B and Plasma (GOT, GPT, Bilirubin) - P | | | | | | | | | |
|---|---|---|---|---|---|---|---|---|---|---|
| | week-0 | | week-2 | | week-4 | | week-6 | | week-8 | |
| | B | P | B | P | B | P | B | P | B | P |
| BDL 200 | | | | | | | | | | |
| BDL 211 | | | | | | | | | | |
| BDL 213 | | | | | | | | | | |
| BDL 326 | | | | | | | | | | |
| BDL 340 | | | | x | | x | | x | | x |
| BDL 341 | | | | x | | x | | x | | x |
| BDL 354 | | x | | | | x | | x | | |
| BDL 355 | | | | | | | | | | |
| BDL 356 | | | | | | | | | | |
| BDL 357 | | | | | | | | | | |
| Sham 358 | | | | | | | | | | |
| Sham 359 | | | | | | | | | | |
| Sham 360 | | | | | | | | | | |
| Sham 361 | | | | | | | | | | |
| Sham 362 | | | | | | | | | | |
| Sham 363 | | | | | | | | | | |
| BDL 364 | | | | | | | | | | |
| BDL 365 | | | | | | | | | | |
| BDL 367 | | | | | | | | | | |
| BDL 368 | | | | | | | | | | |
| BDL 385 | | x | | x | | x | | x | | |
| Sham 388 | | | | | | | | | | |
| BDL 389 | | x | | x | | x | | x | | |
| BDL 404 | x | x | x | x | x | x | | | | |
| BDL 405 | x | x | x | x | | | | | | |
| BDL 406 | x | x | x | x | | | | | | |
| BDL 407 | x | x | x | x | x | x | x | x | | |
| BDL 412 | x | x | x | x | x | x | x | x | | |
| BDL 413 | x | x | x | x | x | x | x | x | | |
| BDL 414 | x | x | x | x | x | x | x | x | x | x |
| BDL 415 | x | x | x | x | x | x | x | x | x | x |
| BDL 466 | x | x | x | x | x | x | x | x | | |
| BDL 467 | x | x | x | x | x | x | x | x | | |
| BDL 468 | x | x | x | x | x | x | x | x | x | x |
| Sham 469 | | | | | | | | | | |
| Sham 470 | | | | | | | | | | |
| BDL 471 | x | x | x | x | x | x | x | x | | |
| BDL 472 | x | x | x | x | x | x | x | x | x | x |
| BDL 474 | x | x | x | x | x | x | x | x | | |
| BDL 475 | x | x | | x | x | x | | | | |
| BDL 476 | | | | | | | | | | |
| BDL 478 | | | | | | | | | | |
| BDL 479 | | | | | | | | | | |
| BDL 483 | | | | | | | | | | |
| BDL 485 | | | | | | | | | | |
| N = 45 | n = 15 | n = 18 | n = 14 | n = 19 | n = 13 | n = 18 | n = 10 | n = 15 | n = 4 | n = 6 |

**Table S2. Detailed report on each rat used in the 1H-MRS experiment.** Number of animals is different per time point due to scanner time limitations and sacrifices for histological measures.

| | ¹H-MRS cerebellum | | | | | ¹H-MRS hippocampus | | | | | ¹H-MRS striatum | | | | |
|---|---|---|---|---|---|---|---|---|---|---|---|---|---|---|---|
| | Week | | | | | week | | | | | week | | | | |
| | 0 | 2 | 4 | 6 | 8 | 0 | 2 | 4 | 6 | 8 | 0 | 2 | 4 | 6 | 8 |
| BDL 200 | x | | x | x | x | x | | x | x | x | | | | | |
| BDL 211 | x | x | x | x | x | x | x | x | x | x | | | | | |
| BDL 213 | x | x | x | x | x | x | x | x | x | x | | | | | |
| BDL 326 | x | | | | x | | | | | | | | | | |
| BDL 340 | x | | x | x | | x | | x | x | | | | | | |
| BDL 341 | x | | x | x | | x | | x | x | | | | | | |
| BDL 354 | x | | x | x | | x | | x | x | | | | | | |
| BDL 355 | | | | x | | | | | | | | | | | |
| BDL 356 | | | | x | | | | | | | | | | | |
| BDL 357 | | | | x | | | | | | | | | | | |
| Sham 358 | | | | | | | | | | | | | | | |
| Sham 359 | | | | | | | | | | | | | | | |
| Sham 360 | | | | | | | | | | | | | | | |
| Sham 361 | | | | | | | | | | | | | | | |
| Sham 362 | | | | | | | | | | | | | | | |
| Sham 363 | | | | | | | | | | | | | | | |
| BDL 364 | | | | | | | | | | | | | | | |
| BDL 365 | | | | | | | | | | | | | | | |
| BDL 367 | | | | | | | | | | | | | | | |
| BDL 368 | | | | | | | | | | | | | | | |
| BDL 385 | x | | x | x | | x | x | | | | | | | | |
| Sham 388 | | | | | | | | | | | | | | | |
| BDL 389 | x | | x | x | | x | | x | x | | | | | | |
| BDL 404 | x | x | x | | | x | x | X | | | x | x | x | | |
| BDL 405 | x | x | | | | x | x | | | | x | x | | | |
| BDL 406 | x | x | | | | x | x | | | | x | x | | | |
| BDL 407 | x | x | x | | | x | x | x | | | x | x | x | | |
| BDL 412 | x | x | x | x | | x | x | x | x | | x | x | x | x | |
| BDL 413 | x | x | x | x | | x | x | x | x | | x | x | x | x | |
| BDL 414 | x | x | x | x | x | x | x | x | x | x | x | x | x | x | x |
| BDL 415 | x | | x | x | x | x | | x | x | x | x | | x | x | x |
| BDL 466 | x | x | x | x | | x | x | x | x | | x | x | x | x | |
| BDL 467 | x | x | x | x | | x | x | x | x | | x | x | x | x | |
| BDL 468 | x | x | x | x | x | x | x | x | x | x | x | x | x | x | x |
| Sham 469 | | | | | | | | | | | | | | | |
| Sham 470 | | | | | | | | | | | | | | | |
| BDL 471 | x | x | x | x | | x | x | x | x | | x | x | x | x | |
| BDL 472 | x | x | x | x | x | x | x | x | x | x | x | x | x | x | x |
| BDL 474 | | x | x | x | | | x | x | x | | | x | x | x | |
| BDL 475 | | | | | | | | | | | | | | | |
| BDL 476 | | | | | | | | | | | | | | | |
| BDL 478 | | | | | | | | | | | | | | | |
| BDL 479 | | | | | | | | | | | | | | | |
| BDL 483 | | | | | | | | | | | | | | | |
| BDL 485 | | | | | | | | | | | | | | | |
| N = 45 | n = 23 | n = 15 | n = 21 | n = 21 | n = 8 | n = 22 | n = 15 | n = 21 | n = 18 | n = 7 | n = 15 | n = 14 | n = 13 | n = 10 | n = 4 |

**Table S3. Detailed report on each rat used for histological measures.**

| | Golgi Cox staining | | GFAP staining | | | Iba-1 staining | | |
|---|---|---|---|---|---|---|---|---|
| | week-8 | control | week-4 | week-8 | control | week-4 | week-8 | control |
| BDL 200 | | | | | | | | |
| BDL 211 | | | | | | | | |
| BDL 213 | | | | | | | | |
| BDL 326 | | | | | | | | |
| BDL 340 | | | | x | | | x | |
| BDL 341 | | | | | | | | |
| BDL 354 | | | | | | | | |
| BDL 355 | x | | | | | | | |
| BDL 356 | x | | | | | | | |
| BDL 357 | x | | | | | | | |
| Sham 358 | | x | | | | | | |
| Sham 359 | | x | | | | | | |
| Sham 360 | | x | | | | | | |
| Sham 361 | | x | | | | | | |
| Sham 362 | | x | | | | | | |
| Sham 363 | | x | | | | | | |
| BDL 364 | x | | | | | | | |
| BDL 365 | x | | | | | | | |
| BDL 367 | x | | | | | | | |
| BDL 368 | x | | | | | | | |
| BDL 385 | | | | | | | | |
| Sham 388 | | | | | x | | | x |
| BDL 389 | | | | | | | | |
| BDL 404 | | | | | | | | |
| BDL 405 | | | | | | | | |
| BDL 406 | | | | | | | | |
| BDL 407 | | | | | | | | |
| BDL 412 | | | | | | | | |
| BDL 413 | | | | | | | | |
| BDL 414 | | | | | | | | |
| BDL 415 | | | | | | | | |
| BDL 466 | | | | | | | | |
| BDL 467 | | | | | | | | |
| BDL 468 | | | | x | | | x | |
| Sham 469 | | | | | x | | | x |
| Sham 470 | | | | | x | | | x |
| BDL 471 | | | | | | | | |
| BDL 472 | | | | x | | | x | |
| BDL 474 | | | | | | | | |
| BDL 475 | | | x | | | x | | |
| BDL 476 | | | x | | | x | | |
| BDL 478 | | | x | | | x | | |
| BDL 479 | | | | | | | | |
| BDL 483 | | | | | | | | |
| BDL 485 | | | | | | | | |
| N = 45 | n = 7 | n = 6 | n = 3 | n = 3 | n = 3 | n = 3 | n = 3 | n = 3 |

## Supplementary Results

BDL animals showed increased blood bilirubin and ammonium from week-2, validating the development of CLD (**Fig.S3**) validating the presence of CLD.

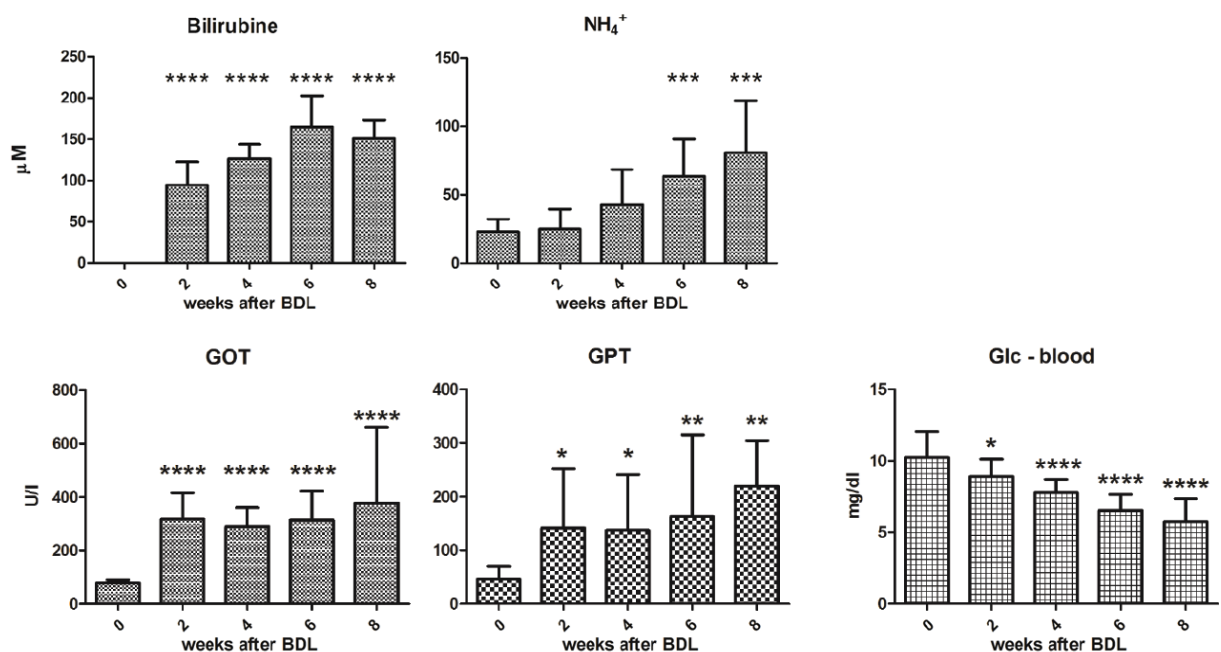

**Fig. S3. Longitudinal changes in total plasma bilirubin, aspartate aminotransferase (AST/GOT) and alanine aminotransferase (ALT/GPT); and blood NH$_4^+$ and glucose (Glc) induced by bile duct ligation.** Bilirubin was non-measurable before BDL. Significance level between week 0 and weeks 2-8: *p<0.05, **p<0.01, ***p<0.001, ****p<0.0001 (1-way ANOVA), mean ± SD.

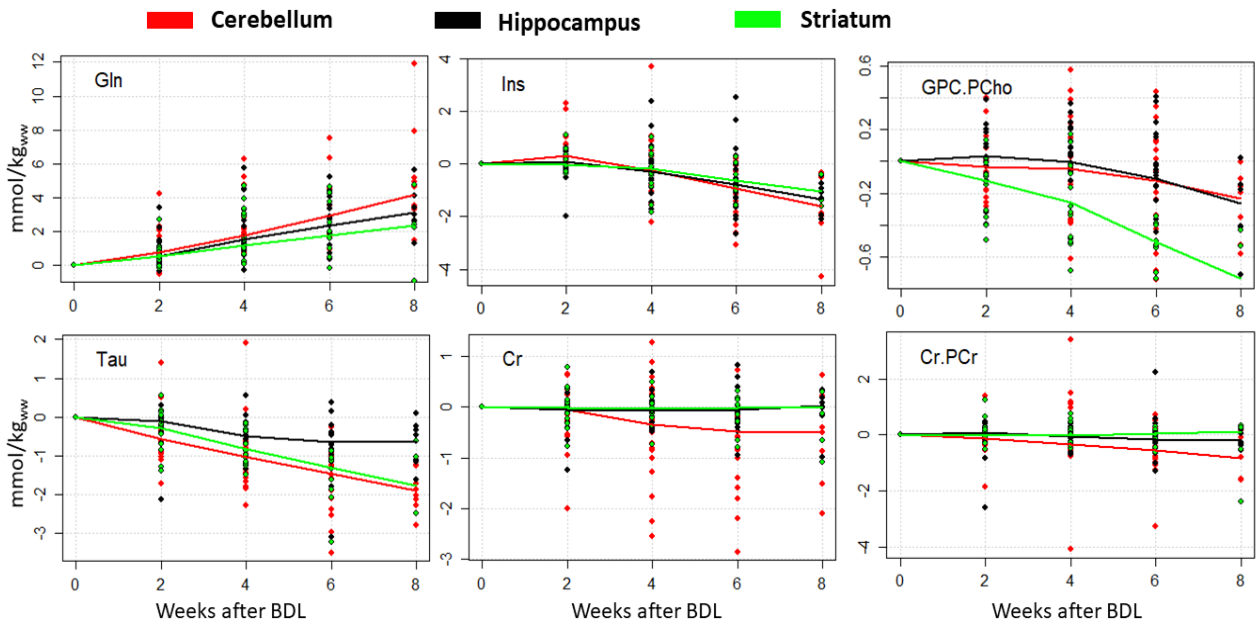

**Fig. S4: Relative changes of glutamine and CNS osmolytes in the three brain regions.** The scatter plots display the relative increase of metabolite concentrations (for Gln, Ins, tCho - GPC.PCho, Tau, Cr and tCr – Cr.PCr) over time during the disease progression.

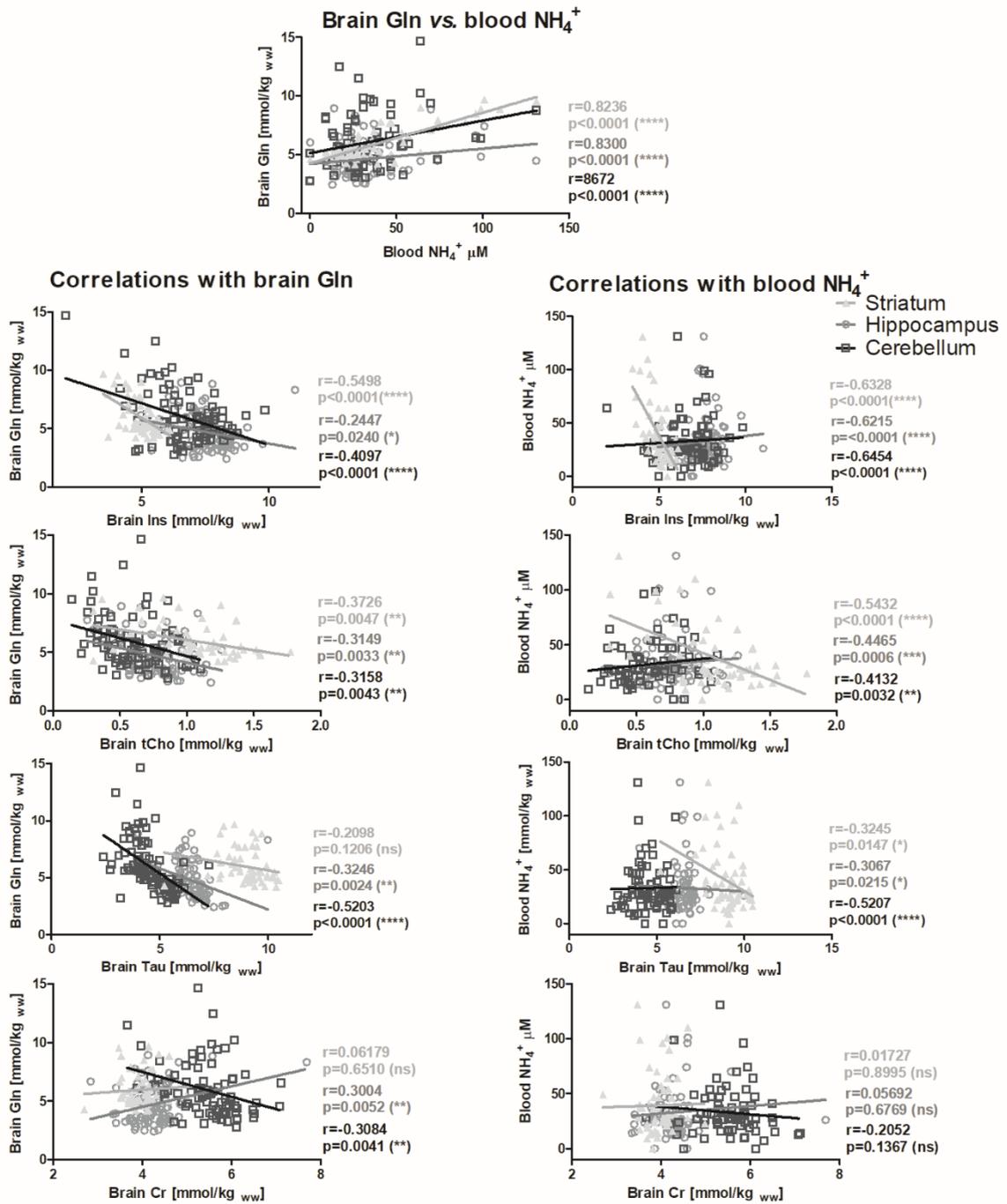

**Fig. S5: Pearson correlations between brain metabolites (Gln, Ins, tCho, Tau and Cr) and blood NH₄⁺** throughout the full course of the study for striatum, hippocampus and cerebellum. The corresponding *p* values are shown along with the Pearson correlation coefficients. Statistical significance: *p<0.05, **p<0.01, ***p<0.001, ****p<0.0001.

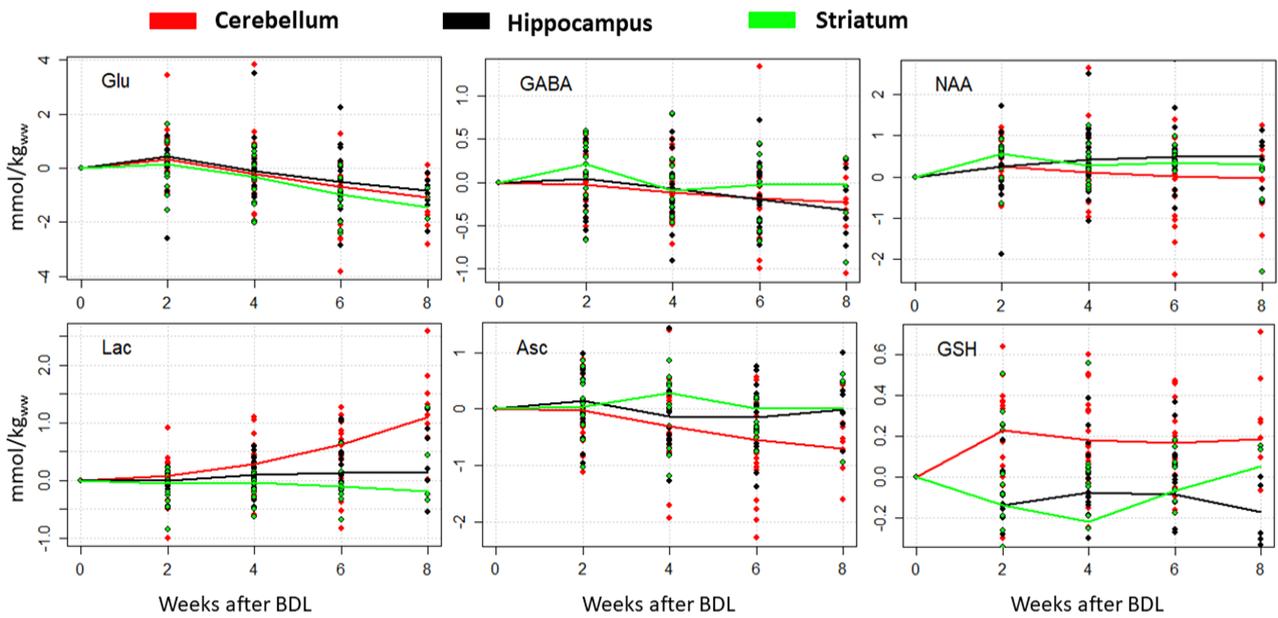

**Fig. S6. Relative changes of other main brain metabolites in the three brain regions**. The scatter plots display relative increase of metabolic concentrations (for Glu, GABA, NAA, Lac, Asc and GSH) over time during the disease progression.

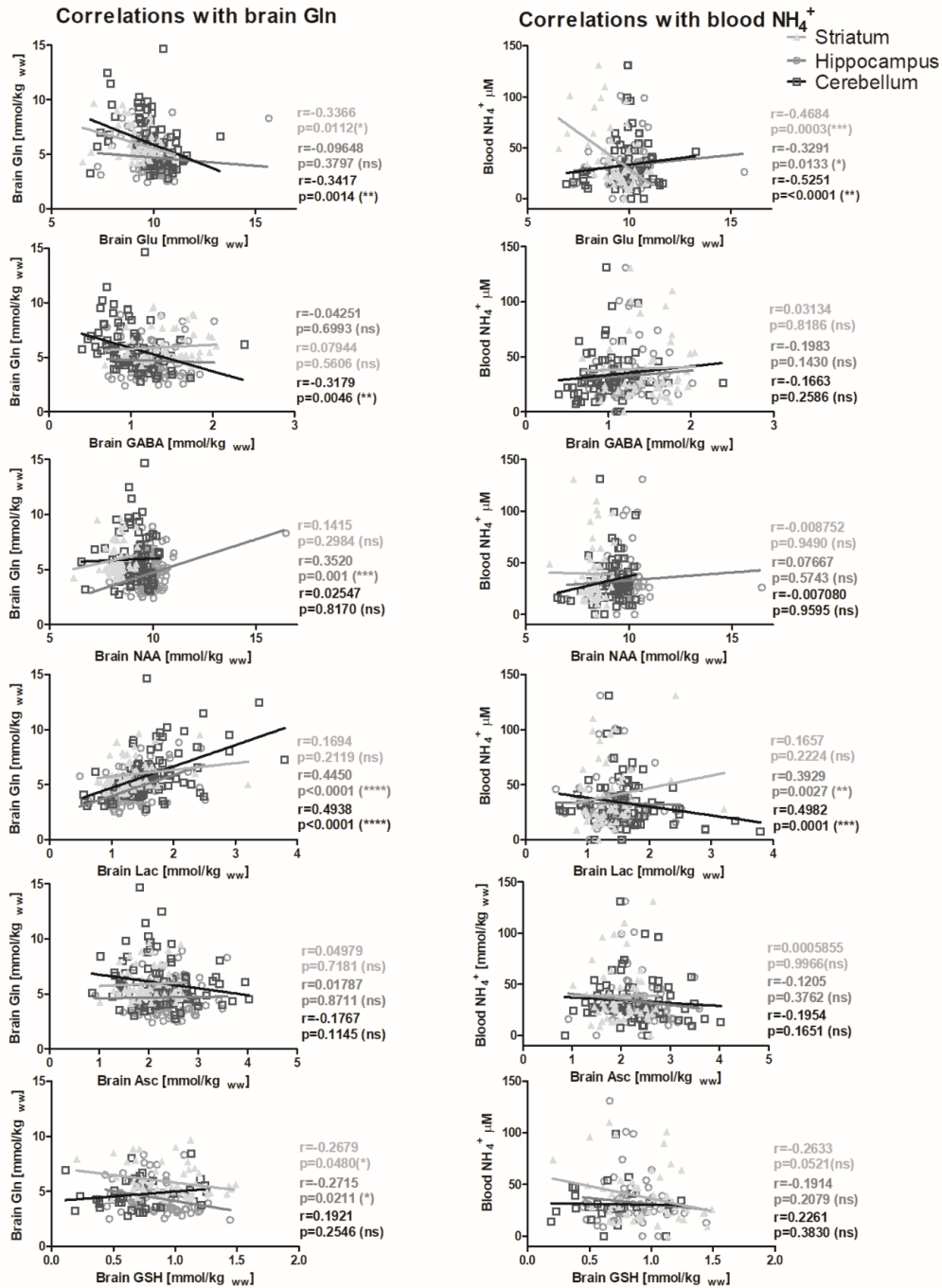

**Fig. S7. Pearson correlations between brain metabolites (for Glu, GABA, NAA, Lac, Asc and GSH) and blood NH₄⁺** throughout the full course of the study for striatum, hippocampus and cerebellum. The corresponding *p* values are shown along with the Pearson correlation coefficients. Statistical significance: *p<0.05, **p<0.01, ***p<0.001, ****p<0.0001.

**Table S4. Results of morphometry measurements of astrocytes (GFAP) and microglia (Iba1) (Sholl analysis) for all the studied brain regions.** Presented as mean ± SD along with the corresponding relative percentage changes. One-way Anova with post-hoc Tukey HSD, *p<0.05, **p<0.01, ***p<0.001, ****p<0.0001, mean ± SD.

| Hippocampus | Astrocytes | | | | | | Microglia | | | | All |
|---|---|---|---|---|---|---|---|---|---|---|---|
| | Number of processes | Processes length [µm] | Processes intersections | | | Astrocytes cells / 0.01 mm² | Number of processes | Processes length [µm] | Perimeter [µm] | Microglia cells / 0.01 mm² | Total nucleus / 0.01 mm² |
| | | | Ring 1 | Ring 2 | Ring 3 | | | | | | |
| Sham | 8.24 ± 1.4 | 34.02 ± 6.5 | 20.88 ± 3.3 | 23 ± 5.3 | 12.5 ± 5.5 | 4.27 ± 0.9 | 4.89 ± 1.13 | 37.2 ± 7.5 | 187.17 ± 25.9 | 1.0 ± 0.41 | 14.51 ± 3.3 |
| BDL w4 | 7.76 ± 1.2 | 29.51 ± 6.8 | 19.71 ± 3.6 | 16.94 ± 2.9 | 6.94 ± 2.7 | 6.29 ± 1.8 | 4.69 ± 0.9 | 44.1 ± 9.43 | 205.93 ± 32 | 1.82 ± 0.44 | 21.07 ± 5.2 |
| BDL w8 | 6.69 ± 1.1 | 22.99 ± 4.9 | 17.85 ± 3.1 | 14.02 ± 4.6 | 3.44 ± 2.7 | 4.61 ± 0.4 | 4.64 ± 1.03 | 39.52 ± 7.69 | 204.13 ± 25.64 | 1.31 ± 0.44 | 18.36 ± 3.2 |
| Relative change [%] | | | | | | | | | | | |
| Sham vs. BDL w4 | - 5.79*** | -13.26**** | - 5.60 | - 26.34**** | - 44.47**** | + 47.48*** | - 4.12 | + 18.57** | + 10.03** | + 39.9** | + 45.26** |
| Sham vs. BDL w8 | - 18.77**** | -32.41**** | - 14.47** | - 39.02**** | - 72.49**** | + 8.09 | - 5.03 | + 6.25** | + 9.07** | + 0.07 | + 26.57* |
| BDL w4 vs. BDL w8 | - 13.78**** | -22.08**** | - 9.4* | - 17.22** | - 50.46**** | - 26.71** | - 0.95 | - 10.39** | - 0.87 | - 27.99** | - 12.87 |

| Cerebellum | Astrocytes | | | | | | Microglia | | | | All |
|---|---|---|---|---|---|---|---|---|---|---|---|
| | Number of processes | Processes length [µm] | Processes intersections | | | Astrocytes cells / 0.01 mm² | Number of processes | Processes length [µm] | Perimeter [µm] | Microglia cells / 0.01 mm² | Total nucleus / 0.01 mm² |
| | | | Ring 1 | Ring 2 | Ring 3 | | | | | | |
| Sham | 11.44 ± 1.6 | 35.66 ± 8.2 | 16.61 ± 2.4 | 14.76 ± 4.7 | 6.82 ± 3.4 | 3.35 ± 0.6 | 4.21 ± 0.97 | 37.38 ± 9.2 | 169.0 ± 18.9 | 1.03 ± 0.36 | 198.24 ± 24.1 |
| BDL w4 | 7.76 ± 1.2 | 29.51 ± 6.8 | 15.25 ± 2.7 | 13.38 ± 3.8 | 6.72 ± 3.9 | 4.86 ± 0.8 | 3.7 ± 0.68 | 47.78 ± 10.8 | 200.25 ± 35.5 | 1.31 ± 0.39 | 303.86 ± 15.8 |
| BDL w8 | 6.69 ± 1.1 | 22.99 ± 4.9 | 13.69 ± 3.1 | 10.72 ± 3.3 | 2.81 ± 1.9 | 4.11 ± 0.6 | 4.74 ± 0.99 | 41.23 ± 8.3 | 205.89 ± 24.9 | 1.06 ± 0.34 | 239.78 ± 34.2 |
| Relative change [%] | | | | | | | | | | | |
| Sham vs. BDL w4 | - 32.17** | -17.26**** | - 8.2** | - 9.4 | - 1.4 | + 45.18*** | - 12.2* | + 27.82** | + 18.49** | + 27.24** | + 53.29*** |
| Sham vs. BDL w8 | - 41.52**** | -35.53**** | - 17.6** | - 27.4**** | - 58.7**** | + 22.72 | + 5.76 | + 9.33** | +17.92** | + 2.53 | + 20.96** |
| BDL w4 vs. BDL w8 | - 13.78**** | -22.08**** | - 10.3* | - 19.9** | - 58.14** | - 15.47** | + 20.87** | - 13.729** | +2.82 | - 19.36* | - 21.09*** |

| Striatum | Astrocytes | | | | | | Microglia | | | | All |
|---|---|---|---|---|---|---|---|---|---|---|---|
| | Number of processes | Processes length [μm] | Processes intersections | | | Astrocytes cells / 0.01 mm² | Number of processes | Processes length [μm] | Perimeter [μm] | Microglia cells / 0.01 mm² | Total nucleus / 0.01 mm² |
| | | | Ring 1 | Ring 2 | Ring 3 | | | | | | |
| Sham | 8.74 ± 1.2 | 37.71 ± 7.9 | 15.81 ± 3.5 | 14.55 ± 5.1 | 6.09 ± 2.3 | 4.18  0.9 | 5.04 ± 0.83 | 35.79 ± 7.67 | 183.14 ± 22.74 | 1.23 ± 0.35 | 32.45 ± 2.9 |
| BDL w4 | 8.31 ± 1.6 | 31.05 ± 5.6 | 16.44 ± 3.1 | 11.89 ± 2.7 | 4.67 ± 2.3 | 5.46 ± 0.7 | 4.69 ± 1.01 | 41.51 ± 9.22 | 201.14 ± 19.36 | 1.7 ± 0.6 | 39.14 ± 5.1 |
| BDL w8 | 7.71 ± 1.2 | 30.14 ± 6.1 | 16.92 ± 4.3 | 12.91 ± 4.5 | 5.17 ± 2.4 | 4.81 ± 0.5 | 4.76 ± 1.06 | 38.07 ± 8.26 | 198.42 ± 25.88 | 1.38 ± 0.42 | 41.6 ± 5.7 |
| Relative change [%] | | | | | | | | | | | |
| Sham vs. BDL w4 | - 4.9 | -17.64*** | + 4 | - 18.3 | - 23.4 | + 30.76*** | - 1.44 | + 15.95** | + 9.83** | + 38.6** | + 20.6*** |
| Sham vs. BDL w8 | - 11.76** | -20.1*** | + 6.9 | - 11.2 | - 15.2 | + 15.2 | - 5.54 | + 6.36* | + 8.35** | + 12.3 | + 28.12*** |
| BDL w4 vs. BDL w8 | - 7.2 | -2.94 | + 2.9 | + 8.6 | + 10.7 | - 11.93* | - 4.15 | - 8.27** | - 1.35 | - 18.96 | + 6.24 |

| Amoeboid microglia | Cells / 0.01 mm² | | |
|---|---|---|---|
| | Cerebellum | Hippocampus | Striatum |
| Sham | ---------------- | ---------------- | ---------------- |
| BDL w4 | 0.33 ± 0.23 | 0.20 ± 0.18 | 0.21 ± 0.30 |
| BDL w8 | 0.67 ± 0.37 | 0.50 ± 0.35 | 0.67 ± 0.58 |
| Relative change [%] | | | |
| BDL w4 vs. BDL w8 | 102** | 150* | 212* |

**Table S5. Results of the Golgi cox staining of hippocampal and cerebellar neurons.** Changes in neuronal soma, apical and basal dendrites length is presented as mean ± SD with the corresponding relative percentage changes. One-way Anova with post-hoc Tukey HSD, *p<0.05, **p<0.01, ***p<0.001, ****p<0.0001, mean ± SD.

| *Hippocampus CA1* | *Neuronal soma [µm²]* | *Spines / 10µm dendrites length* | |
|---|---|---|---|
| | | *Apical* | *Basal* |
| *Sham* | 227.47 ± 60.81 | 9.65 ± 2.09 | 9.97 ± 1.93 |
| *BDL w6* | 385.74 ± 105.43 | 6.41 ± 1.67 | 6.06 ± 1.49 |
| *Relative change [%]* | | | |
| *Sham vs. BDL w8* | +69.57*** | - 33.64*** | - 39.20*** |

| *Hippocampus DG* | *Neuronal soma [µm²]* | *Apical Spines / 10µm dendrites length* |
|---|---|---|
| *Sham* | 137.50 ± 36.21 | 9.56  1.54 |
| *BDL w6* | 175.13 ± 49.07 | 6.45 ± 1.73 |
| *Relative change [%]* | | |
| *Sham vs. BDL w8* | + 27.36*** | - 32.51*** |

| *Cerebellum Purkinje Cells* | *Neuronal soma [µm²]* | *Apical Spines / 10µm dendrites length* |
|---|---|---|
| *Sham* | 488.28 ± 111.27 | 12.89 ± 2.14 |
| *BDL w6* | -347.42 ± 87.86 | 9.19 ± 1.81 |
| *Relative change [%]* | | |
| *Sham vs. BDL w8* | -28.84*** | - 28.67**** |

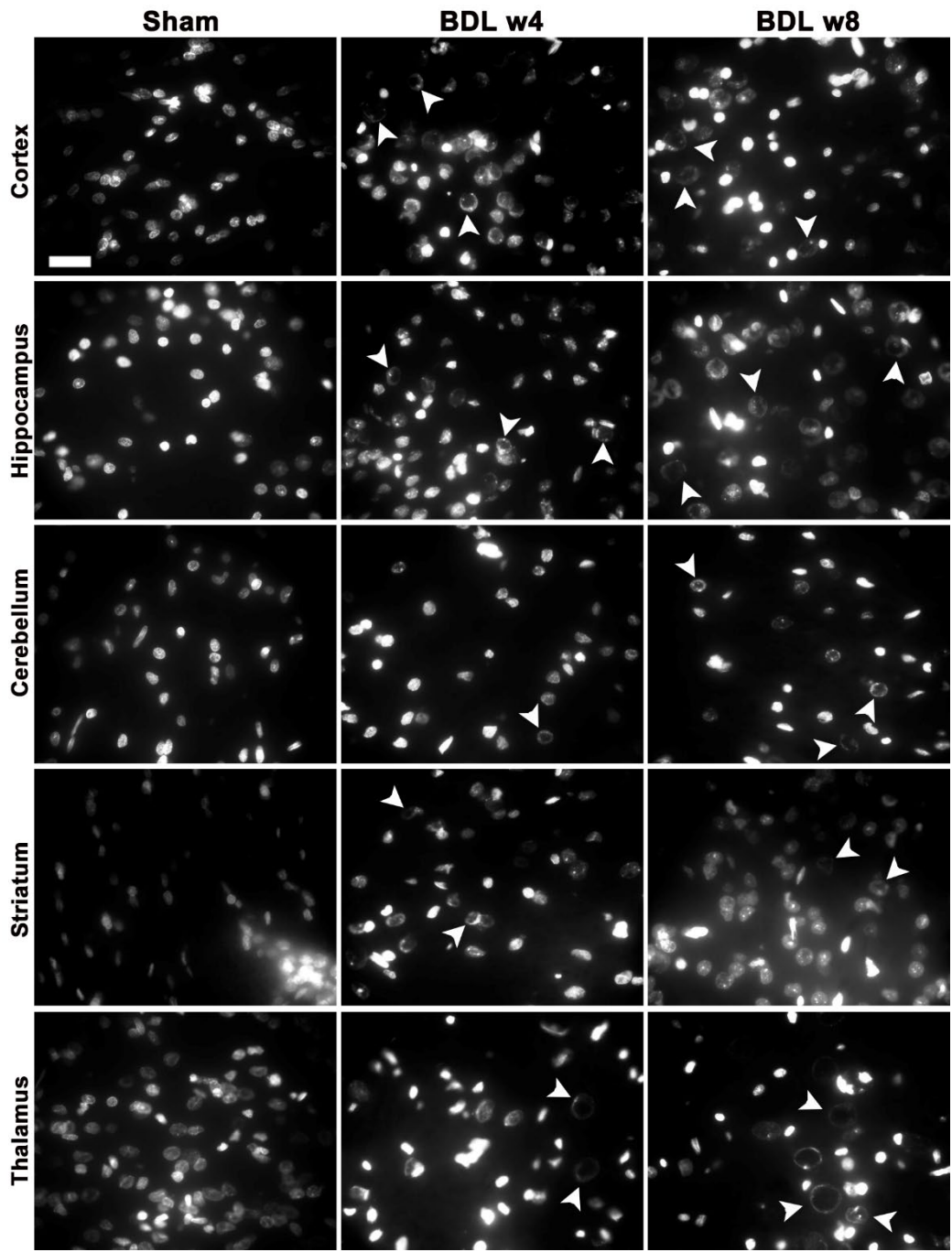

**Fig. S8. Representative micrograph of DAPI stained brain sections illustrating type II astrocytes** (arrowheads). Type II astrocytes show nuclear enlargement, chromatin migration and clearing (scale bar = 25μm).

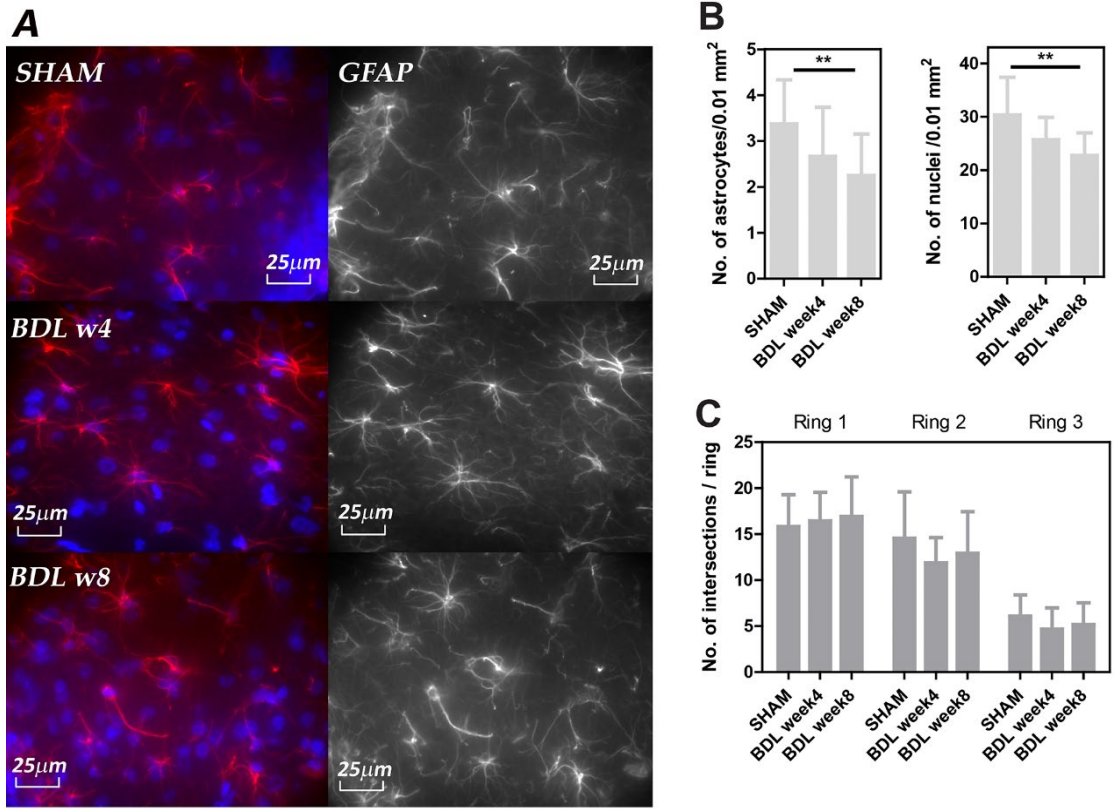

**Fig. S9. Astrocytes Sholl analysis of striatum.** **(A)** Brain sections from Sham and BDLs rats at 4- and 8-weeks post-op were stained with GFAP (red) and DAPI (blue). **(B)** Astrocytes and total nuclei number, astrocytes processes number and process lengths quantification. **(C)** The number of intersections across the given rings starting from the cell body's center - plotted according to subregions (1-3) (Sholl analysis, **Fig. S1**). One-way Anova with post-hoc Tukey HSD, *p<0.05, **p<0.01, ***p<0.001, ****p<0.0001, mean ± SD.

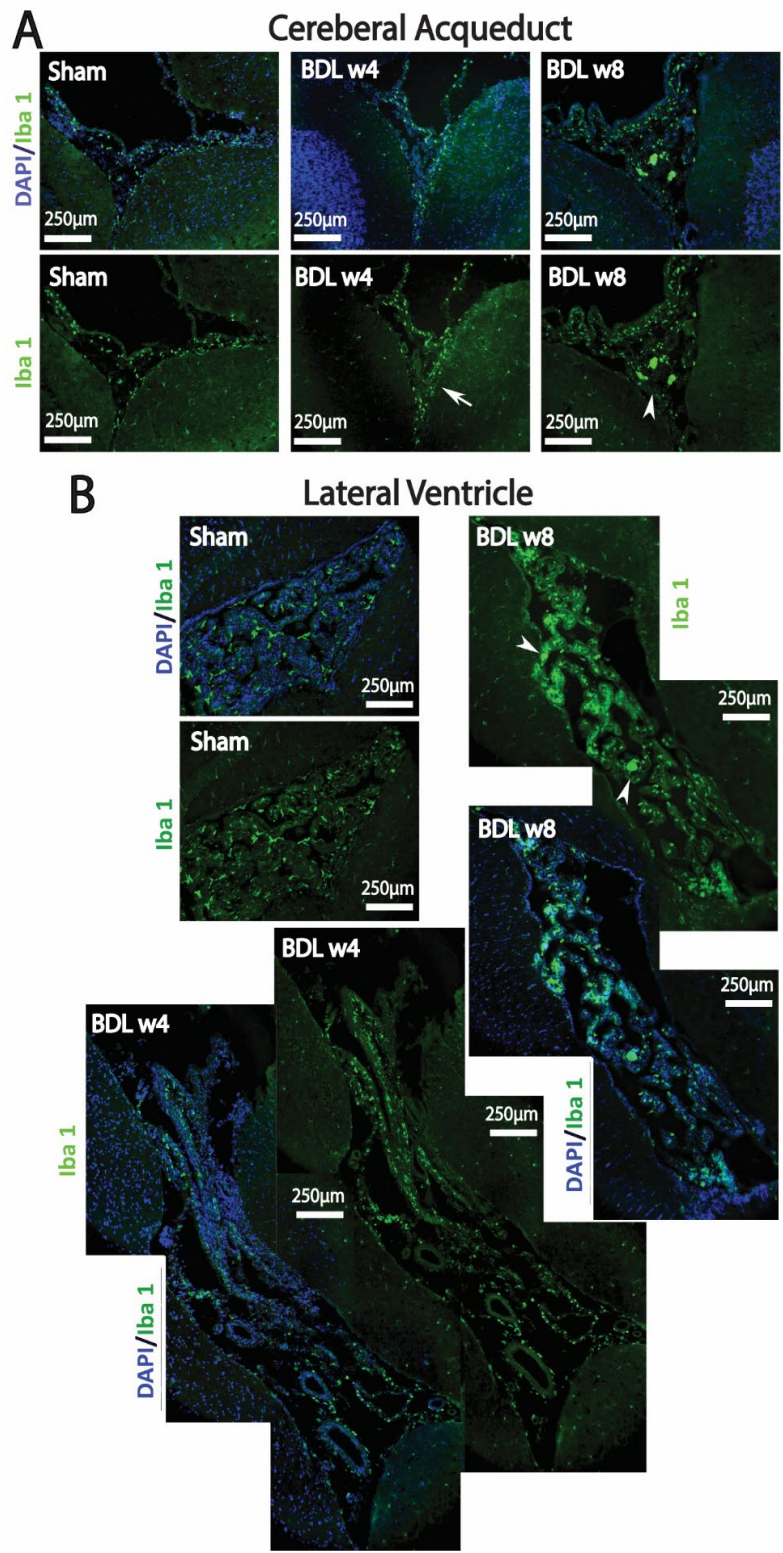

**Fig. S10. Representative micrographs of histological sections of ventricles.** Anti-Iba1 (green) and DAPI (blue) staining of **(A)** mesencephalic cerebral aqueduct and **(B)** telencephalic lateral ventricle. Activated macrophages accumulation and cloth like formation. White arrow - increased number of macrophages, white arrowhead – macrophages clots like formation.

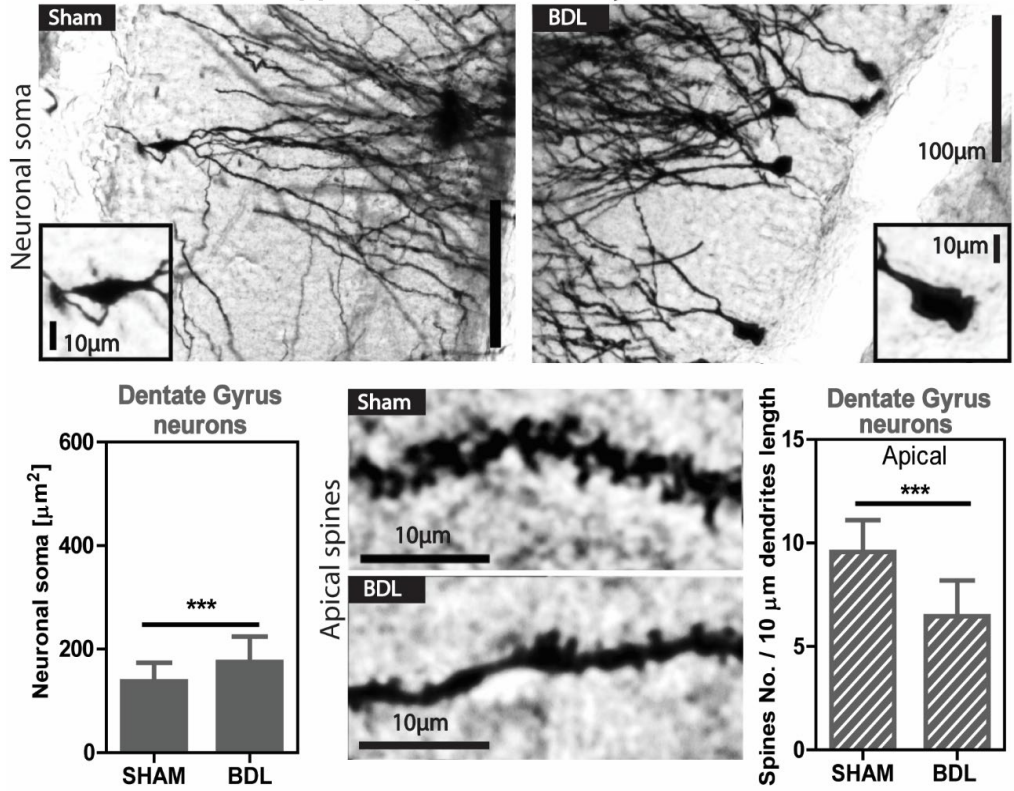

**Fig. S11. Representative micrographs of histological sections of Golgi-Cox staining and neuronal morphology analysis of hippocampal Dentate Gyrus neurons.** One-way Anova with post-hoc Tukey HSD: *p<0.05, **p<0.01, ***p<0.001, ****p<0.0001, mean ± SD.